\documentclass[pra,onecolumn,superscriptaddress,aps]{revtex4}
\usepackage{amsfonts}
\usepackage{amssymb,latexsym,amsmath}
\usepackage{graphicx}
\usepackage{times}
\usepackage[caption=false]{subfig}
\usepackage[usenames]{color}
\usepackage{float}

\begin{document}

\title{Rogue Waves and Periodic Solutions of a Nonlocal Nonlinear
  Schr{\"o}dinger Model}

\author{C. B. Ward}

\affiliation{Department of Mathematics and Statistics, University of
  Massachusetts, Amherst, Massachusetts 01003-4515, USA}

\author{P. G. Kevrekidis}
\affiliation{Department of Mathematics and Statistics, University of
Massachusetts, Amherst, Massachusetts 01003-4515, USA}
\affiliation{Mathematical Institute, University of Oxford, Oxford, OX2
  6GG, UK}

\author{T.~P. Horikis}
\affiliation{Department of Mathematics, University of Ioannina, Ioannina 45110, Greece}

\author{D.~J. Frantzeskakis}
\affiliation{Department of Physics, National and Kapodistrian University of Athens,
Panepistimiopolis, Zografos, Athens 15784, Greece}

\begin{abstract}
In the present work, a nonlocal nonlinear Schr{\"o}dinger (NLS) model is studied by means of
a recent technique that identifies solutions of partial differential equations, by considering
them as fixed points in {\it space-time}. This methodology allows to perform a continuation of
well-known solutions of the local NLS model to the nonlocal case. Four different examples
of this type are presented, namely (a) the rogue wave in the form of the Peregrine soliton,
(b) the generalization thereof in the form of the Kuznetsov-Ma breather, as well as
two spatio-temporally periodic solutions in the form of elliptic functions. Importantly,
all four waveforms can be continued in intervals of the parameter controlling the nonlocality
of the model. The first two can be continued in a narrower interval, while the periodic ones
can be extended to arbitrary nonlocalities and, in fact, present an intriguing bifurcation whereby
they merge with (only) spatially periodic structures. The results suggest the generic relevance
of rogue waves and related structures, as well as periodic solutions, in nonlocal NLS models.
\end{abstract}

\maketitle

\section{Introduction}

The study of dispersive media exhibiting a nonlocal nonlinear response is a subject that is
enjoying increasing attention over the past few years~\cite{t28,assanto1,assanto2}.
This is mainly due to the fact that relevant models and their solutions, especially of
the nonlinear Schr{\"o}dinger (NLS) variety, emerge in a wide range of physical contexts.
These range from thermal optical media~\cite{t28,t27} and nematic liquid
crystals~\cite{assanto1,assanto2,t25} and from plasmas~\cite{t29,t30} to water waves~\cite{hf1,hf2}
and dipolar Bose-Einsein condensates \cite{bec1,bec2}. It is, thus, naturally of interest to explore the
different types of solitary wave solutions that may arise in nonlocal NLS systems
and how nonlocality may alter the properties and behavior of media with a local nonlinear
response~\cite{neshev,krol2,krol3}.

Another topic of wide interest during the past decade has been the study of rogue (freak) waves,
i.e., coherent structures of large amplitude that appear out of nowhere and disappear without
a trace~\cite{wandt}. The study of such waves has been explored mainly in
hydrodynamics~\cite{hydro,hydro2,hydro3}, but also in numerous other areas. These include, but are
not limited to, nonlinear optics~\cite{opt1,opt2,opt3,opt4,opt5,laser}, superfluid
helium~\cite{He}, as well as plasmas~\cite{plasma}. These multifaceted experimental studies have,
in turn, triggered a wide range of theoretical explorations which by now have been summarized in a
series of reviews \cite{onorato,solli2,yan_rev,Akhmediev2016,Chen2017,Mihalache2017,MalomedMihalache2019},
but also importantly in a series of books on this research theme~\cite{k2a,k2b,k2c,k2d}.

In the present work, we combine these two cutting edge themes, by exploring rogue waves and related
coherent structures in nonlocal media motivated by the above optical, liquid crystal and water wave
applications. It should be noted that a search of the literature of rogue waves in nonlocal NLS
models will yield a number of results, such as, e.g.~\cite{jyang}. However, these concern a
mathematically motivated (via {\cal PT}-symmetry and related considerations) nonlocal variant of
the NLS~\cite{AM}. There is a paucity of results concerning rogue waves of physically relevant
nonlocal NLS models, as e.g., the one considered in Ref.~\cite{ha3}. However, this is rather
understandable given the non-integrability of such models and the distinct lack of tools for
tackling rogue waves beyond the integrable limit and its associated techniques, such as the
inverse scattering method.

Here, we aim to provide a number of results regarding the model of the nonlocal NLS that is of wide
relevance to applications. Our approach leverages a recently proposed technique reported in Ref.~\cite{ward}.
In this work, it was recognized that it is difficult to identify rogue waves (contrary to what is the case
with solitons), due to their non-stationary nature in time. However, considering them in {\it
space-time}, i.e., treating time as a spatial direction, one realizes that rogue waves are
localized solutions in that setting. Thus, one can ---in principle--- construct fixed point methods
(based, e.g., on a  conjugate gradient-based variant of the Newton method~\cite{ward}) that will
converge to such wave structures, and identify them as numerically exact solutions. A key advantage
of such a methodology is that it does not hinge in any critical way on integrability and, indeed,
starting from the integrable limit it can be used in a variety of non-integrable settings, such as
the nonlocal one that we consider here. It is this tool that will permit us to converge to the rogue wave
(in the form of the Peregrine soliton) for a range of parameter values of the nonlocality parameter,
referred to as $\nu$ below in our model. In addition to the prototypical Peregrine structure,
we will also seek its periodic ---in the propagation variable--- generalization, namely
the Kuznetsov-Ma (KM) breather. Both of these will be found to be possible to continue within
a certain interval of the nonlocality parameter (roughly up to $\nu=0.2$). Apart from these
structures, we will also consider and identify additional states that our analysis can
reveal in the nonlocal system, namely periodic states in space-time (starting from the elliptic
function local limit). We will see that these can be continued for essentially arbitrary $\nu$, yet
they feature a bifurcation becoming stationary (or independent of the propagation direction) beyond
a certain critical threshold.

Our presentation will be structured as follows. In section II, we will present the solutions of
interest in the local limit of the regular NLS model (i.e., for $\nu=0$). In section III, we will
present the numerical extensions of the solutions via the above method to the nonlocal case of
finite $\nu$. Finally, in section IV, we will summarize our findings and present our
conclusions.

\section{The Model and Its Local Limit Solutions}

We consider a nonlocal variant of the NLS model, which, in dimensionless form, can be expressed  as~\cite{ha3,assanto1,assanto2}:
\begin{gather}
i\frac{\partial u}{\partial t} + \frac{1}{2} \frac{\partial^2
  u}{\partial x^2} + \theta u - \mu u =0,
\label{neq1}
  \\[.2cm]
\nu \frac{\partial^2 \theta}{\partial x^2} - 2q\theta = -2 |u|^2.
\label{neq2}
\end{gather}
where $u=u(x,t)$, $\theta=\theta(x,t)$, and, in the context of nematic liquid crystals, $t$ plays
the role of the propagation coordinate. The dependent variable $u$ is the complex valued,
slowly varying envelope of the optical (electric) field, and $\theta$ is the optically induced
deviation of the director angle. The nonlocality parameter $\nu$ measures the strength of the
response of the nematic in space, with a highly nonlocal response corresponding to $\nu$ large.
Notice that in the nonlocal regime with $\nu$ large, the optically induced
rotation $\theta$ is small. The parameter $q$ is related to the square of the applied static
field which pre-tilts the nematic dielectric, while $\mu$ plays the role of the propagation constant.
Obviously, inclusion of the term $\mu u$ in Eq.~(\ref{neq1}) (which is normally omitted) offers a constant background where
certain types of solutions (such as the Peregrine soliton ---see below) can exist, and is trivially removed via a constant phase transformation. Hereafter, we set $q=1$.

The system (\ref{neq1})-(\ref{neq2}) can be considered as a single integro-differential equation
provided Fourier transforms are utilized. This will not limit our set of solutions however for we
will make only numerical use of the transform pair. As such a periodic domain is used to integrate
the system in $x$ and nondecaying functions will not generate problems with the transform not
converging. Thus, using the Fourier Transform we can rewrite Eq.~(\ref{neq2}) as
\begin{equation}
\theta = \mathcal{F}^{-1}\left[\frac{\mathcal{F}[2|u|^2]}{\nu k^2+2}\right].
\end{equation}
Plugging this into Eq.~(\ref{neq1}) yields a nonlocal equation in $u$ alone and will be the
equation with which we work in our numerical computations hereafter. Note that when $\nu=0$ we
recover the standard NLS equation.

%The system (\ref{neq1})-(\ref{neq2}) can be considered as an
%integro-differential single equation
%provided Fourier transforms are utilized.
%%This will not limit our set of solutions that we will consider
%%here, as we will make only numerical use of the transform pair. As
%%such,
%In line with the numerical use of the Fourier tranforms below,
%a periodic domain will be used
%to integrate the system in $x$.
%%and nondecaying functions will not generate problems with the
%%Fourier transform not converging. Thus, using the Fourier Transform
%Then, we can rewrite Eq.~(\ref{neq2}) as:
%%
%\begin{equation}
%\theta = \mathcal{F}^{-1}\left[\frac{\mathcal{F}[2|u|^2]}{\nu k^2+2}\right],
%\end{equation}
%%
%where $\mathcal{F}$ (and $\mathcal{F}^{-1}$) denotes the Fourier (and inverse Fourier) transform.
%Plugging this into Eq.~(\ref{neq1}) yields a nonlocal equation in $u$ alone and will be the
%equation with which we work in our numerical computations hereafter. Note that when $\nu=0$ we
%recover the standard NLS equation.

%%%% Pano: maybe explain the numerical scheme here?

%We will continue four different solutions of the NLS starting from
%$\nu=0$.
The four solutions that will be of interest hereafter are analytically
available in the NLS limit.
The first one is the famous Peregrine soliton~\cite{H_Peregrine};
for $\mu=1$, this solution reads:
\begin{eqnarray}
u(x,t)= 1-\frac{2(1+4it)}{1+4x^2 +4t^2}.
\label{neq3}
\end{eqnarray}
This structure has been the subject of numerous recent experimental observations in
hydrodynamics~\cite{hydro}, nonlinear optics~\cite{opt2}, plasmas~\cite{plasma}, and so on.

The second one is the periodic generalization of the Peregrine soliton in the
evolution direction, namely the so-called Kuznetsov-Ma (KM) breather given by~\cite{kuz,ma}:
\begin{eqnarray}
%\mu=1\\
u(x,t)= 1-\frac{2(b^2-1) \cos(2b\sqrt{b^2-1}t)+i2b\sqrt{b^2-1}
  \sin(2b\sqrt{b^2-1}t)}{b
  \cosh(2\sqrt{b^2-1}x)-\cos(2b\sqrt{b^2-1}t)},
  \label{neq4}
\end{eqnarray}
where $b$ is an arbitrary parameter (with $b>1$). It is worthwhile
to note that this solution has been experimentally observed as well~\cite{opt3}.

We now consider some spatio-temporally periodic solutions stemming
from the classic work of~\cite{akh}, both of them for simplicity given
for $\mu=1/2$.
The first is a doubly periodic solution given by:
\begin{eqnarray}
%\mu=1/2\\
u(x,t)& =&\frac{\kappa}{2} \; \frac{A(x) {\rm cn}(t/2)+i\sqrt{1+\kappa} \;
  {\rm sn}(t,\kappa)}{\sqrt{1+\kappa}-A(x){\rm dn}(t/2,\kappa)}, \\
A(x) &=& {\rm cd}\bigg(\sqrt{\frac{1+\kappa}{2}}x, \sqrt{\frac{1-\kappa}{1+\kappa}} \bigg),
  \label{neq5}
\end{eqnarray}
where $\kappa$ is an arbitrary parameter (with $0<\kappa<1$).
The second is another doubly periodic solution given by:
\begin{eqnarray}
%\mu=1/2\\
u(x,t) =\frac{\sqrt{\frac{\kappa}{1+\kappa}}
  {\rm cn}\big(\frac{x}{\sqrt{\kappa}},\sqrt{\frac{1-\kappa}{2}}\big)
  {\rm dn}\big(\frac{t}{2\kappa},\kappa\big) +i\kappa \; {\rm sn} \big(\frac{t}{2\kappa} ,
  \kappa\big)}{\kappa\sqrt{2}\big[1 - \sqrt{\frac{\kappa}{1+\kappa}} {\rm cn}\big(\frac{x}{\sqrt{\kappa}},
  \sqrt{\frac{1-\kappa}{2}} \big) {\rm cn} \big(\frac{t}{2\kappa}, \kappa\big)\big]},
  \label{neq6}
\end{eqnarray}
where $\kappa$ is again an arbitrary parameter (with $0<\kappa<1$).

%%%%%%%%%%%%%%%%%%FIGURES%%%%%%%%%%%%%%%%%%%%%%%%
\section{Continuation to the Nonlocal Case}

\subsection{Peregrine Soliton}

\begin{figure}[]
% Fig 1
\center
\subfloat[$\nu=0$]{\includegraphics[width=.24\textwidth]{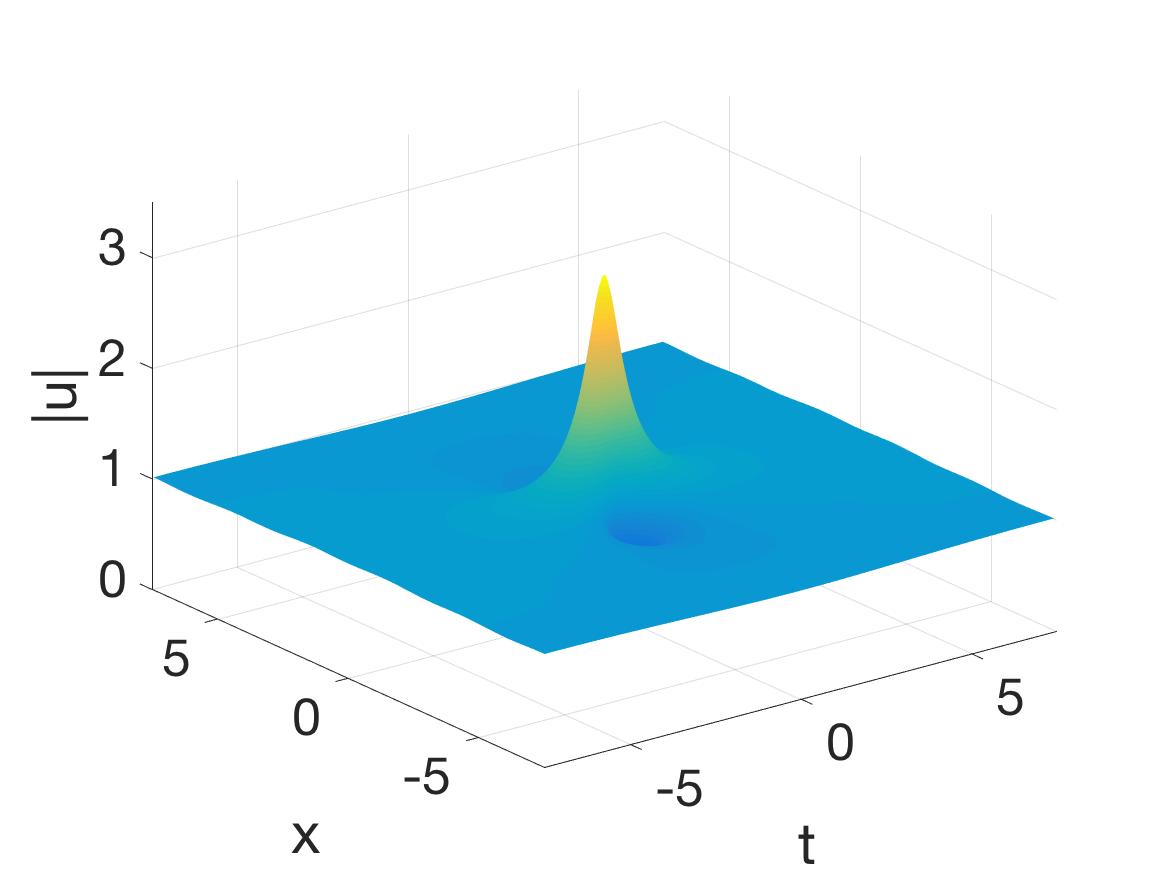}}
\subfloat[$\nu=0.05$]{\includegraphics[width=.24\textwidth]{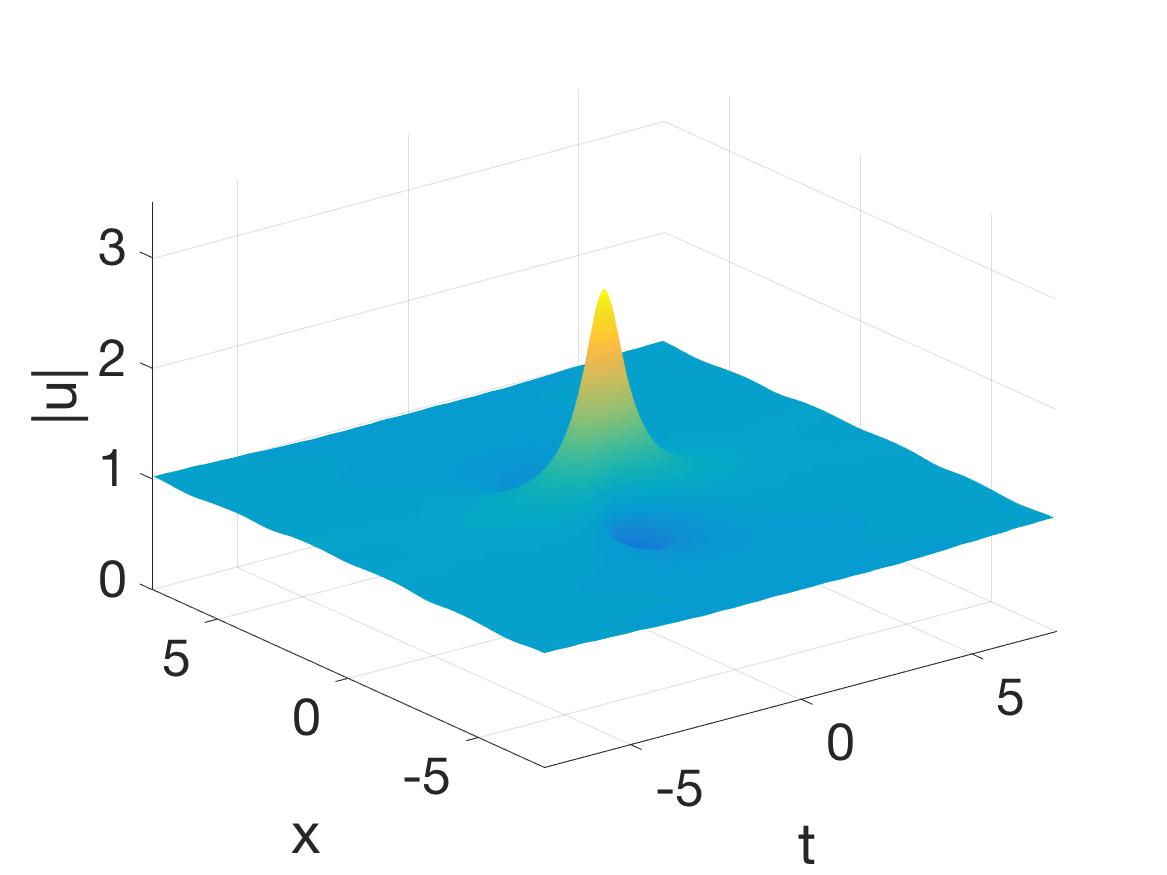}}
%\subfloat[$\nu=0.1$]{\includegraphics[width=.3\textwidth]{Fig4c.jpg}}
\subfloat[$\nu=0.15$]{\includegraphics[width=.24\textwidth]{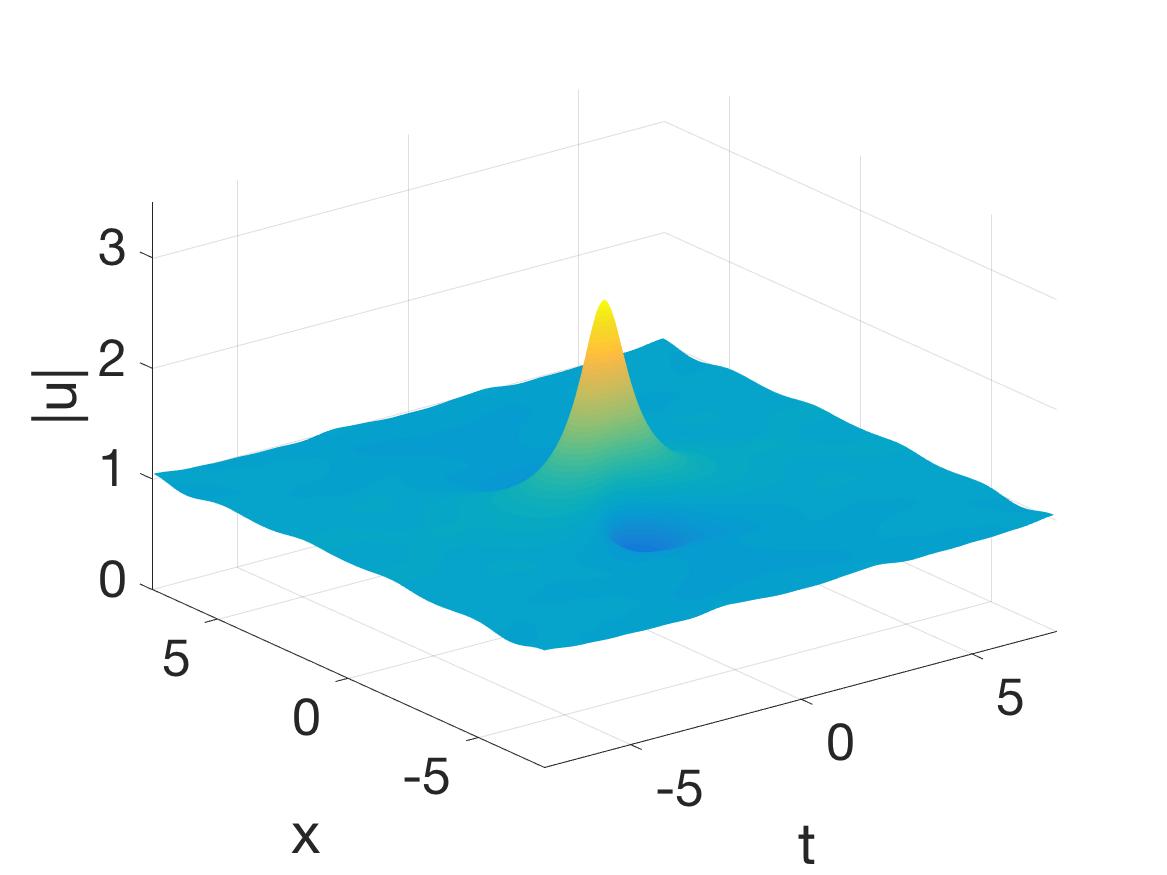}}
\subfloat[$\nu=0.2$]{\includegraphics[width=.24\textwidth]{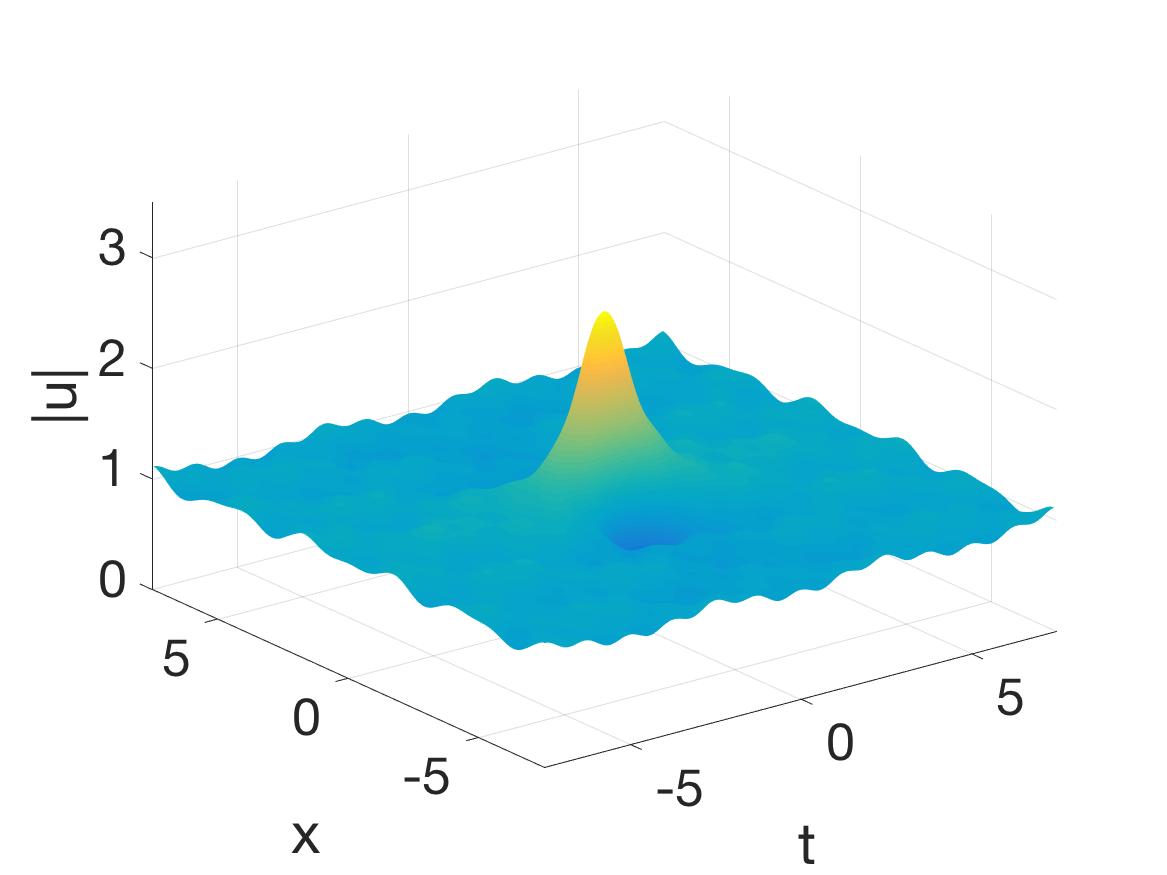}}

\caption{Continuation over increasing $\nu$ values of the Peregrine
  rogue wave solution of the NLS. The absolute value of the
  wavefunction is shown as a function of space $x$ and time $t$, once
  the
  iterative identification of the solution converges.}
\label{fig:Fig8}
\end{figure}

\begin{figure}[]
% Fig 2
\center
\subfloat[$\nu=0$]{\includegraphics[width=.24\textwidth]{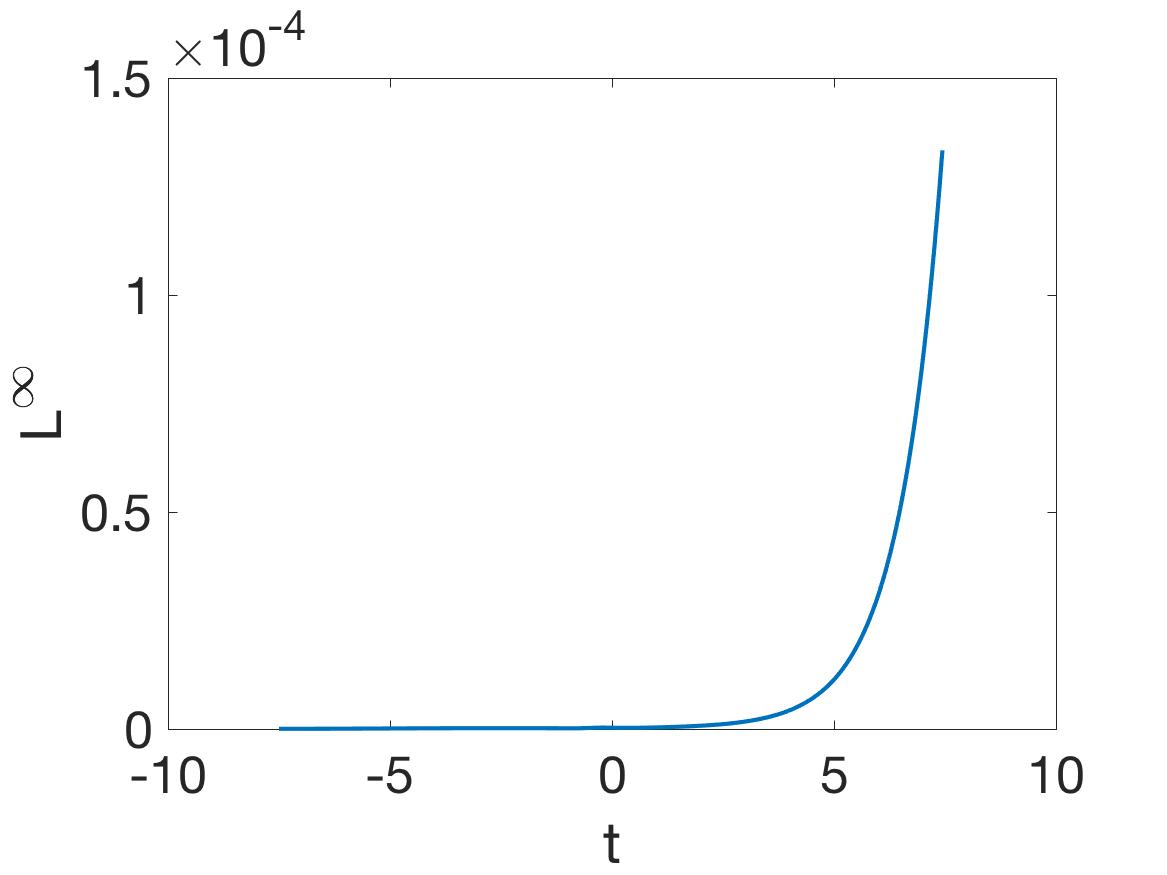}}
%\subfloat[$\nu=0.05$]{\includegraphics[width=.27\textwidth]{Fig12b.jpg}}
\subfloat[$\nu=0.1$]{\includegraphics[width=.24\textwidth]{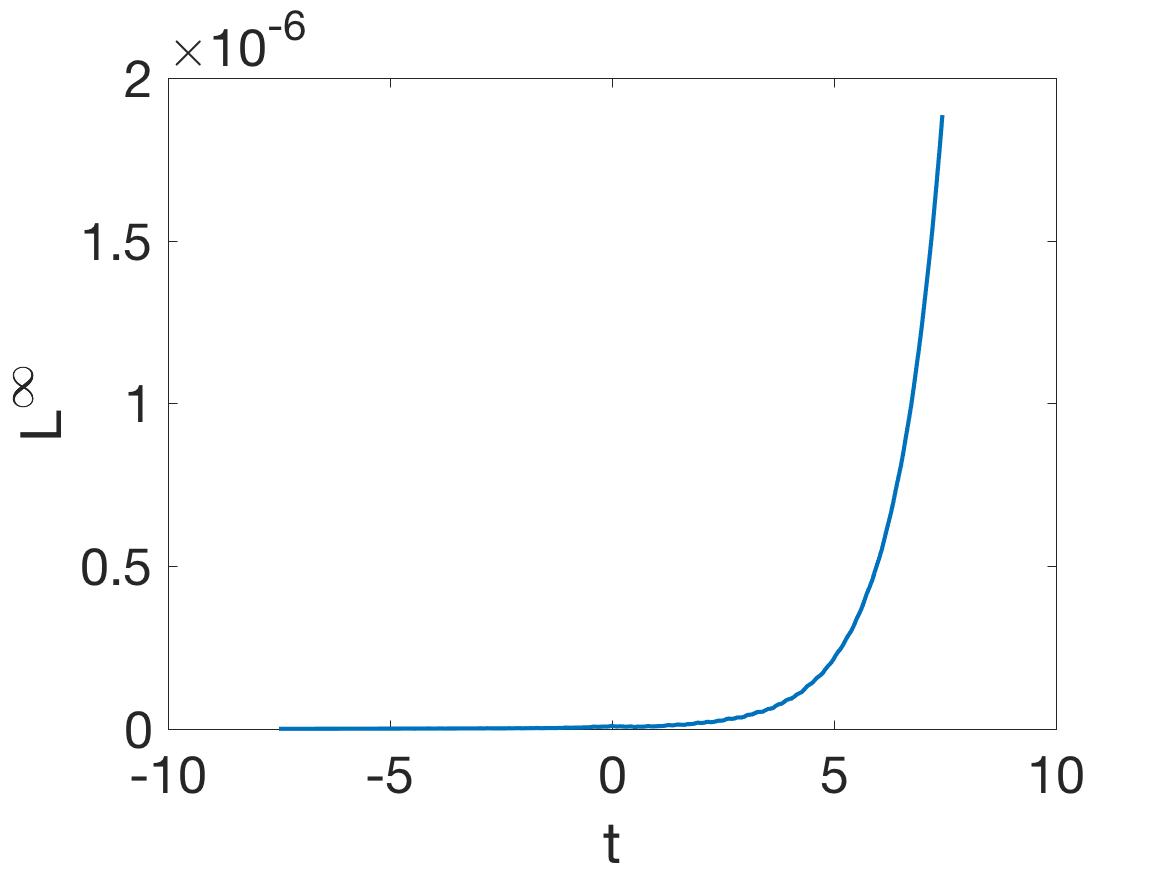}}
\subfloat[$\nu=0.15$]{\includegraphics[width=.24\textwidth]{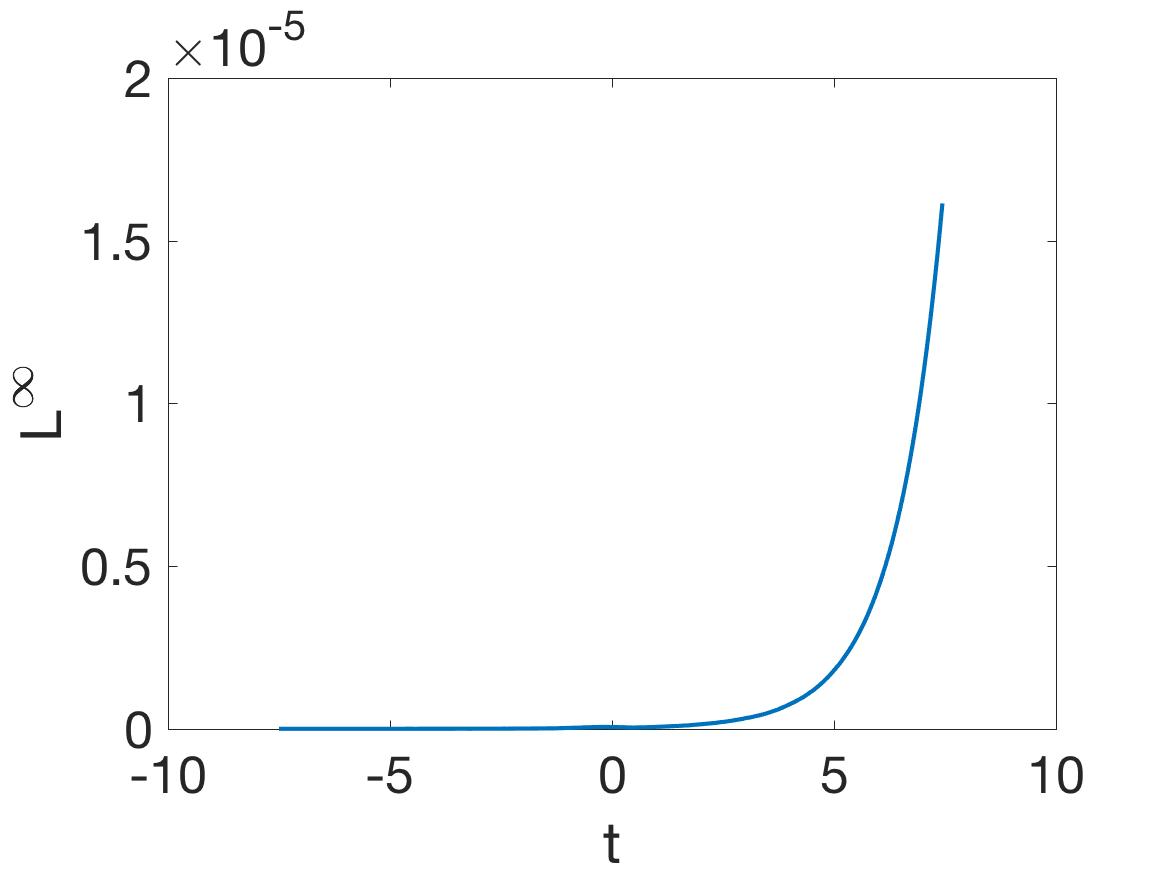}}
\subfloat[$\nu=0.2$]{\includegraphics[width=.24\textwidth]{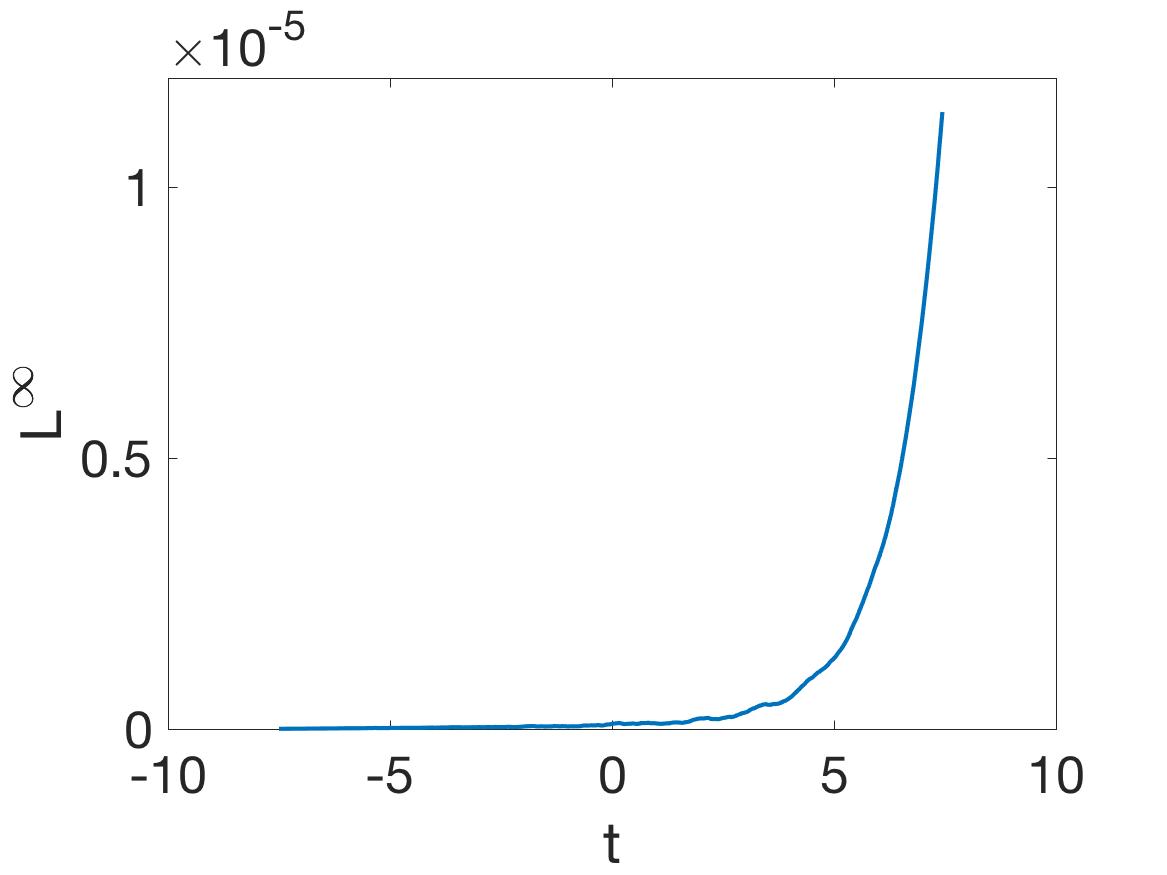}}

\caption{$L^\infty$ error (i.e., norm of the difference) between the
  numerical ETDRK4 solution of Eqs.~(\ref{neq1})-(\ref{neq2}) and the
  converged Peregrine soliton waveform obtained by the Newton-CG method for
  the different case examples of nonlocality parameter $\nu$ shown in Fig.~\ref{fig:Fig8}.}
\label{fig:Fig12}
\end{figure}

The results associated with the Peregrine soliton are given in
Figs.~\ref{fig:Fig8} and~\ref{fig:Fig12}. The first one, shows the profile
of the Peregrine soliton as the value of the nonlocality parameter $\nu$ is increased.
Perhaps the most important finding in itself is that this structure
can still be obtained as a numerically exact solution beyond the integrable
limit, and in the nonlocal case of $\nu \neq 0$. Structurally, it can be observed
that the solution acquires a certain ``undulation'', as $\nu$ is
increased, that becomes progressively more pronounced. It is important to also
highlight here that the solution is identified with periodic boundary
conditions in both space and time (recall that time is treated as a space
variable so a periodicity is imposed on that as well).
The continuation scheme is unable to go past $\nu=0.2$, even when considering different
domain sizes. Nevertheless, we do not detect a bifurcation at this point,
hence it is unclear whether this is a trait of the solution or a by-product of
the particular numerical method. We believe that the latter is true.

Figure~\ref{fig:Fig12}, in turn, is a dynamical illustration of the
accuracy of convergence of our solution. Here, what is done is that
we select the ``initial condition'' of our converged Peregrine waveform at $t=-7.5$
and feed it into an integrator of the full nonlocal problem at
different values of $\nu$. The forward propagation of
Eqs.~(\ref{neq1})-(\ref{neq2}) is performed using the exponential
time-differencing 4th-order Runge-Kutta (ETDRK4) method of Ref.~\cite{Kassam}.
In the figure it is observed that the measured $L^{\infty}$ error
(i.e., norm of the difference) between the converged solution
and the numerically propagated one remains very small (i.e., of
O$(10^{-4})$ to O$(10^{-5})$) throughout the propagation. We should
factor in here the numerical instability of the solution, which
eventually leads to growth (well past the formation and disappearance
of the Peregrine soliton). Nevertheless, for all practical purposes, the
converged state accurately captures the appearance and disappearance of the
coherent structure.

\subsection{Kuznetsov-Ma Breather}

\begin{figure}[]
% Fig 3
\center
\subfloat[$\nu=0$]{\includegraphics[width=.24\textwidth]{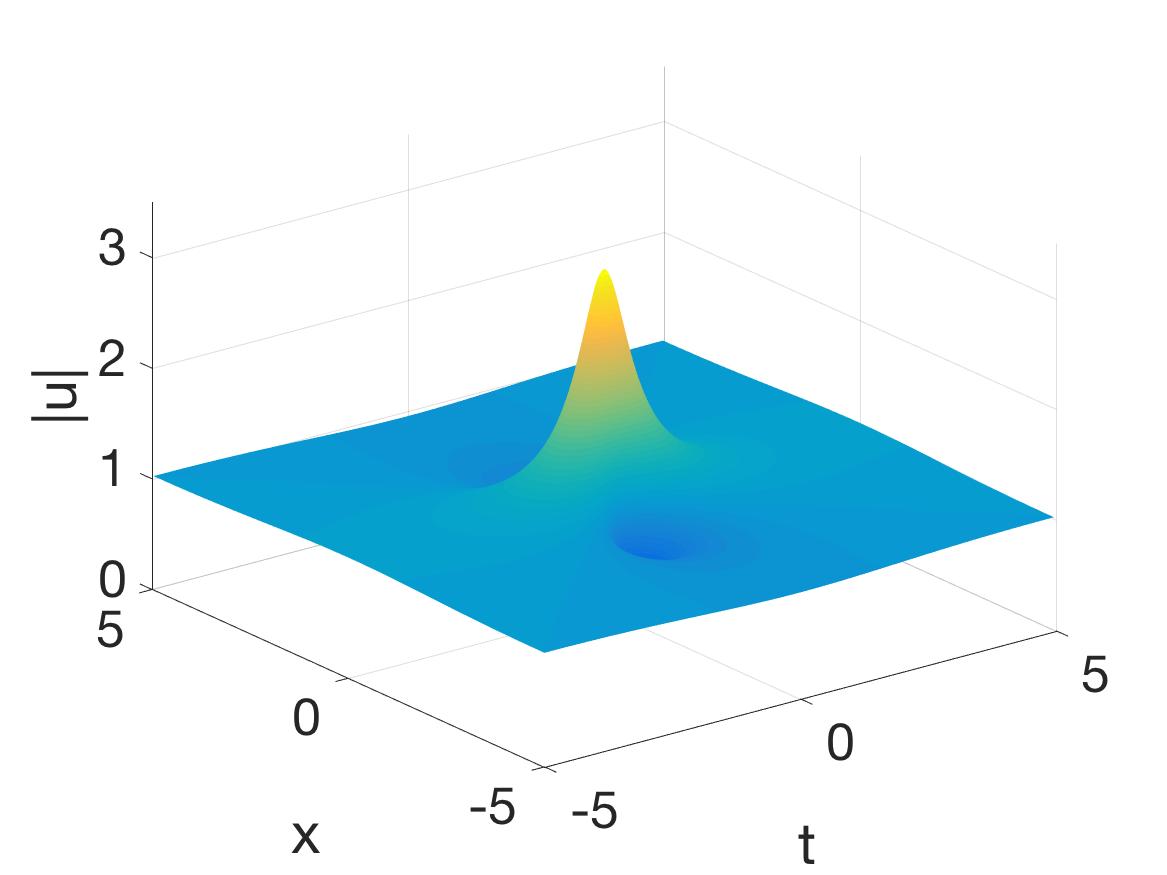}}
%\subfloat[$\nu=0.05$]{\includegraphics[width=.3\textwidth]{Fig3b.jpg}}
\subfloat[$\nu=0.1$]{\includegraphics[width=.24\textwidth]{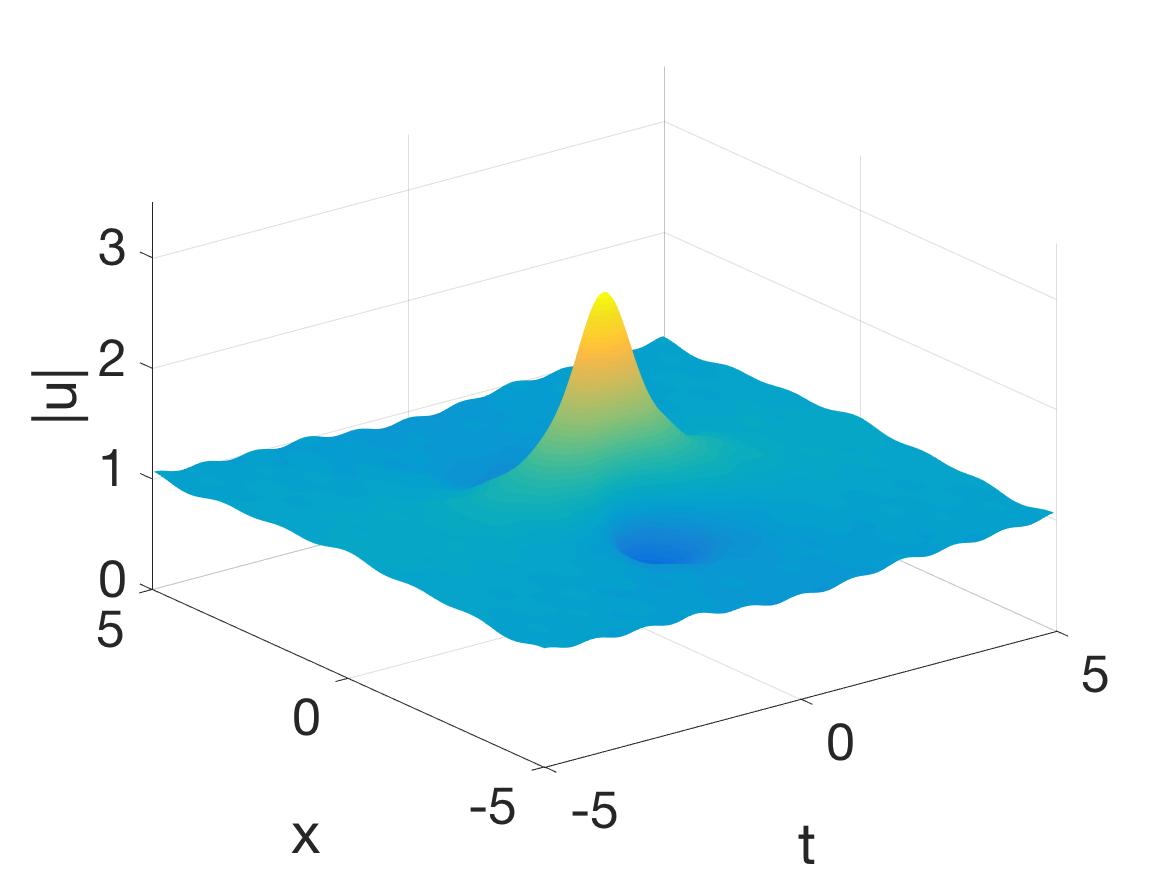}}
\subfloat[$\nu=0.15$]{\includegraphics[width=.24\textwidth]{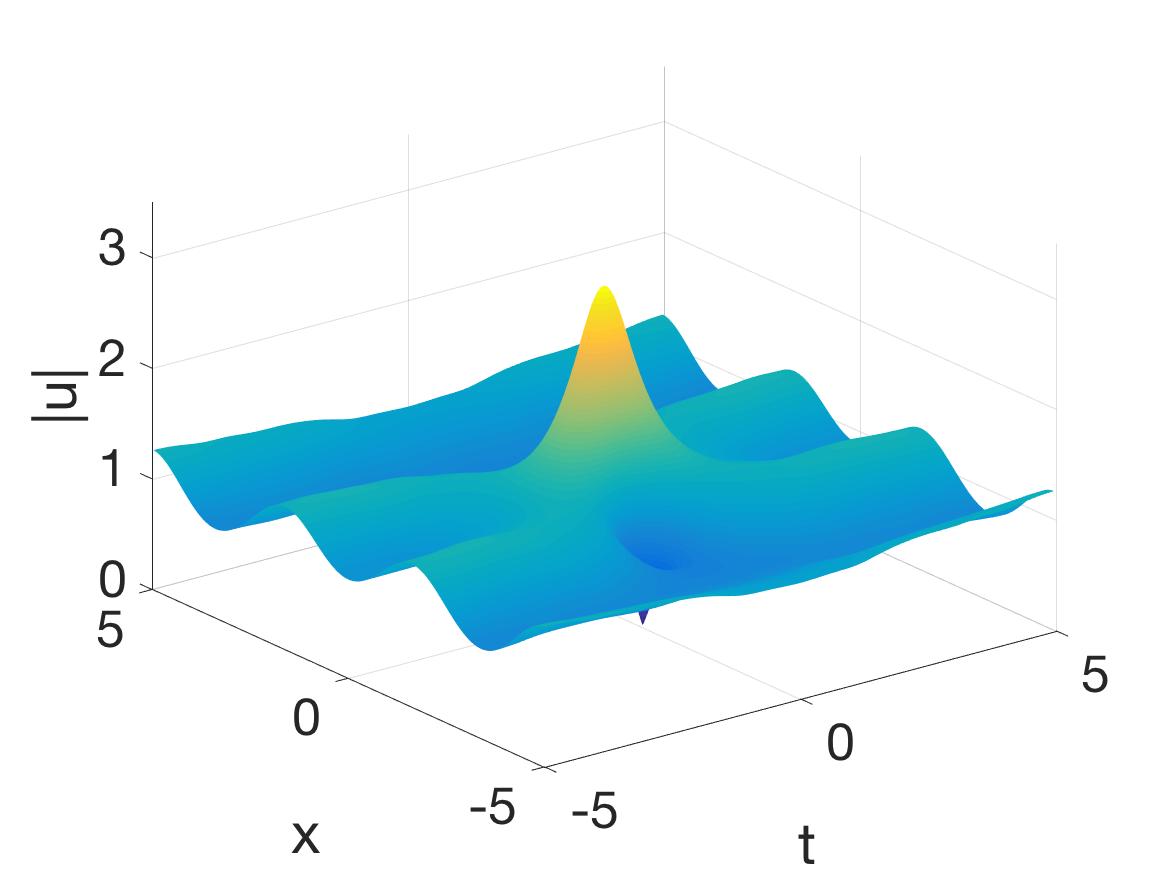}}
\subfloat[$\nu=0.2$]{\includegraphics[width=.24\textwidth]{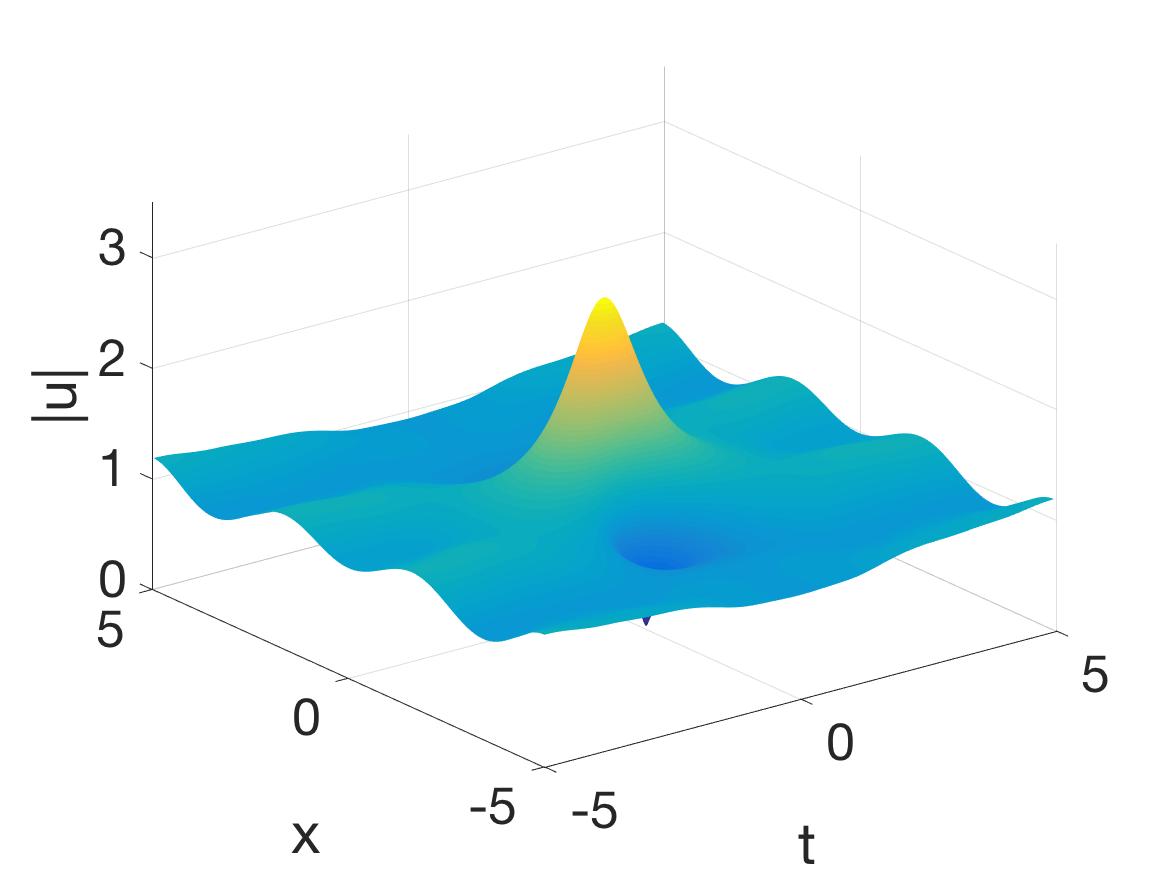}}

\caption{Continuation over increasing $\nu$ values of the KM breather
  solution of the NLS. Note that the solution seems to transition into
  a breather on top of a periodic background as the nonlocality
  parameter is increased.}
\label{fig:Fig7}
\end{figure}

\begin{figure}[]
% Fig 4
\center
\subfloat[$\nu=0$]{\includegraphics[width=.24\textwidth]{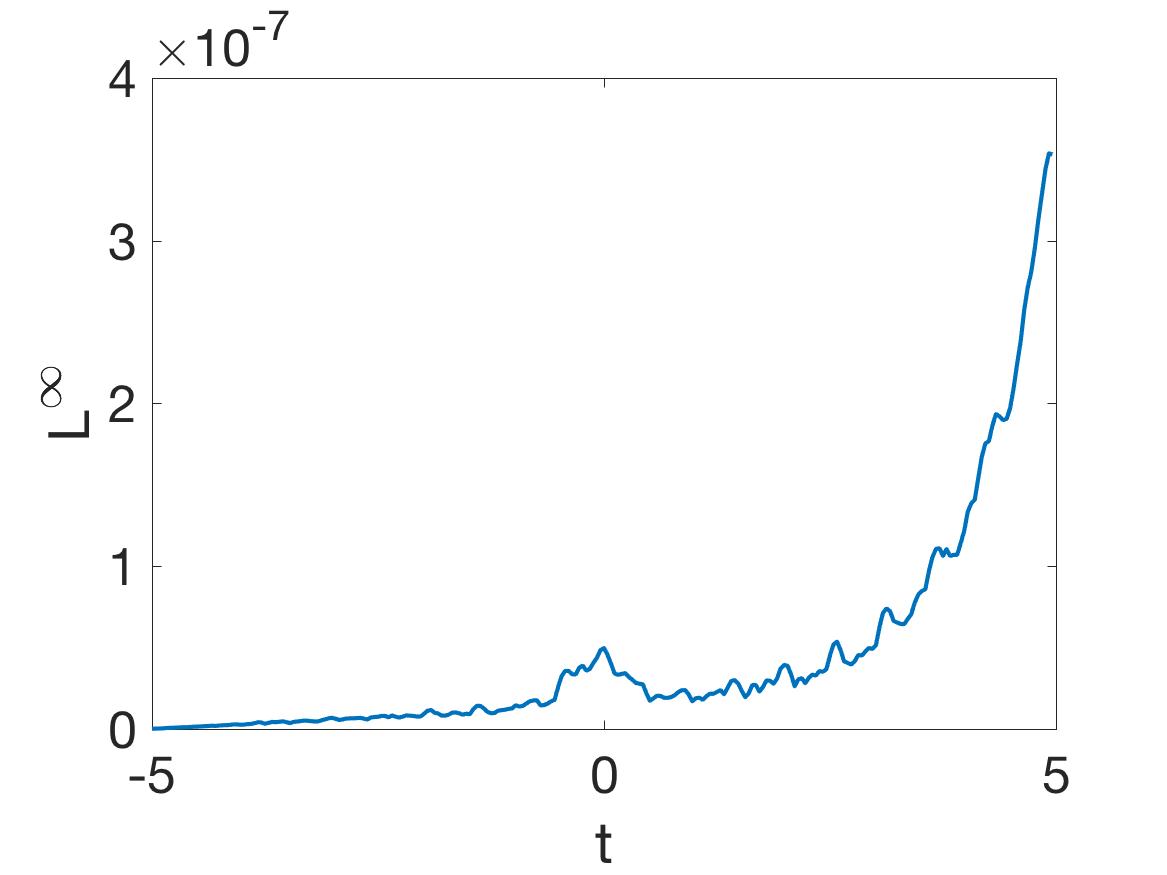}}
%\subfloat[$\nu=0.05$]{\includegraphics[width=.27\textwidth]{Fig11b.jpg}}
\subfloat[$\nu=0.1$]{\includegraphics[width=.24\textwidth]{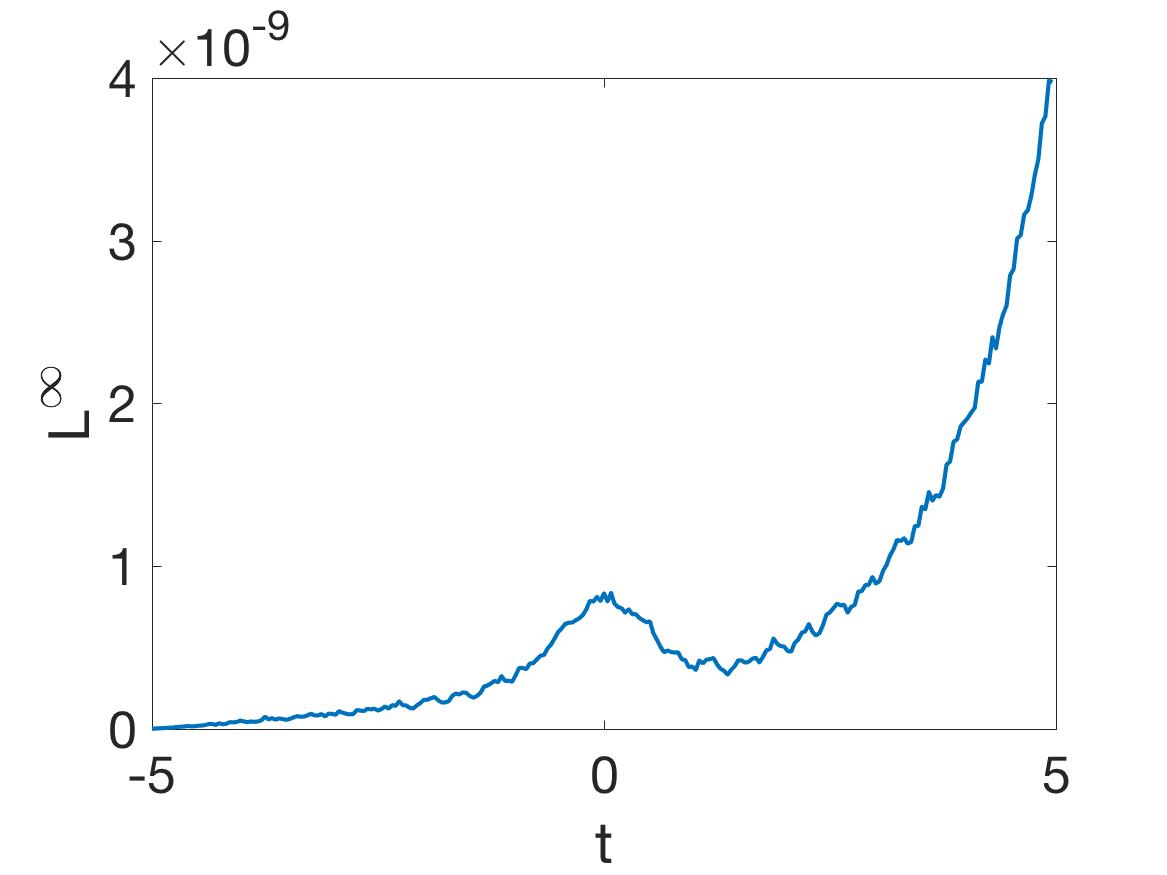}}
\subfloat[$\nu=0.15$]{\includegraphics[width=.24\textwidth]{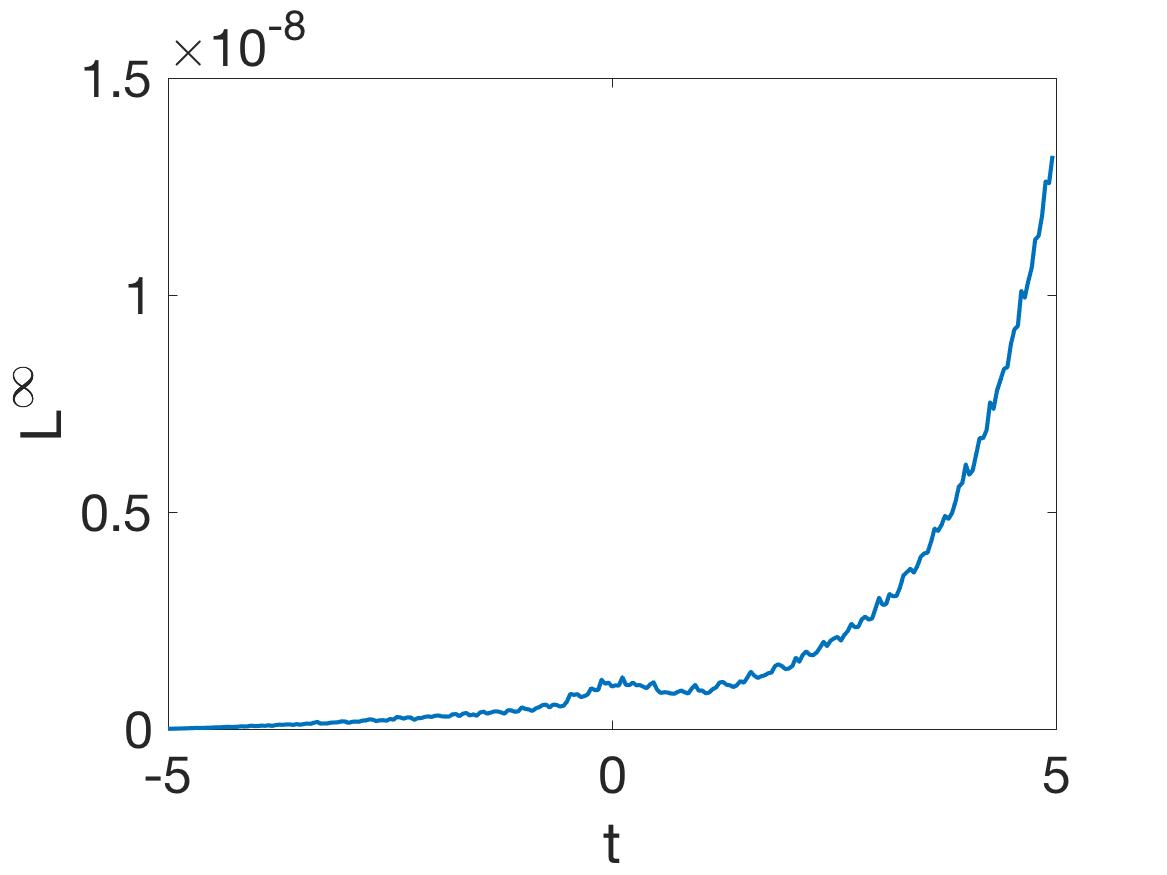}}
\subfloat[$\nu=0.2$]{\includegraphics[width=.24\textwidth]{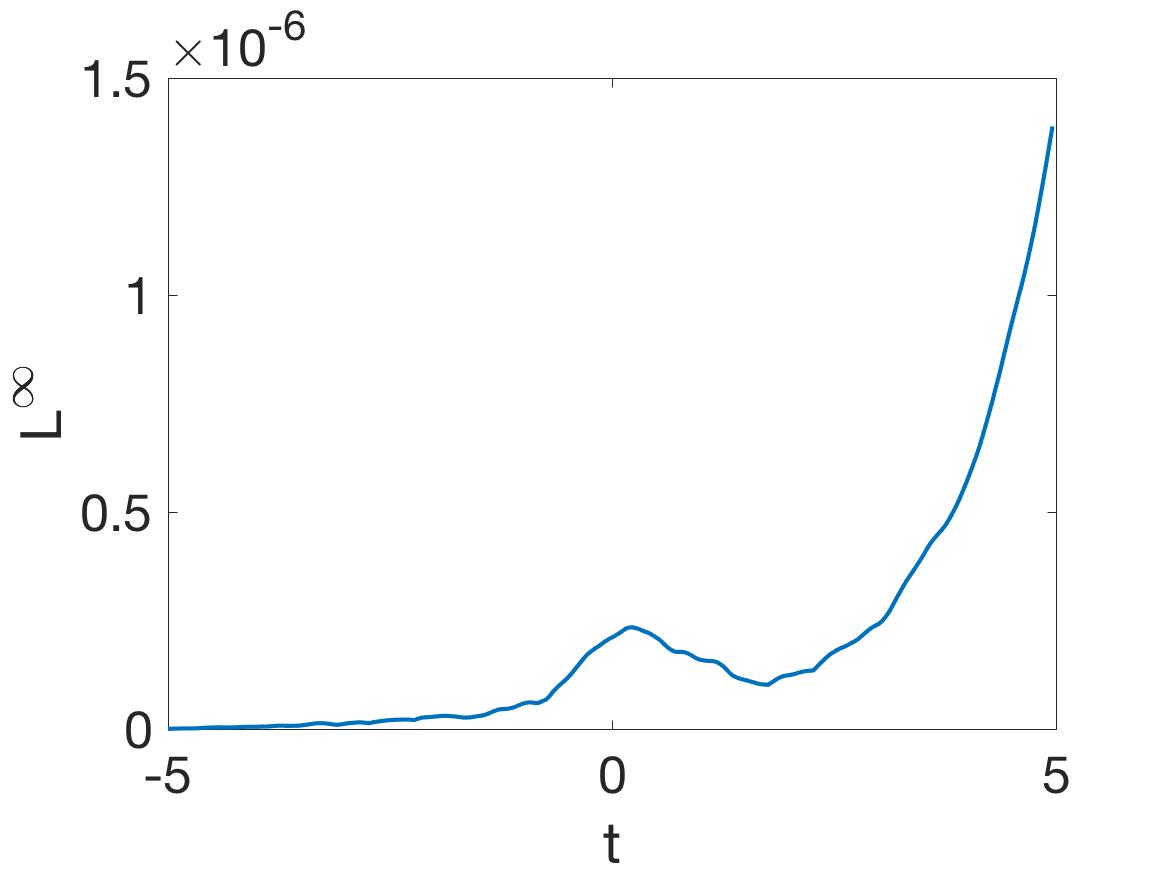}}

\caption{$L^\infty$ error between the solution obtained by the
  Newton-CG method (as shown in Fig.~\ref{fig:Fig7}) vs the solution
  obtained via the ETDRK4  method when the KM breather profile is
  seeded into the latter as an initial condition.}
\label{fig:Fig11}
\end{figure}

In this subsection, we present similar diagnostics for the KM breather
waveform of Eq.~(\ref{neq4}). It is interesting that, here, the
background modulation becomes more transparent and arises in the clear form of
a progressively more intense (as the nonlocality parameter $\nu$
increases) periodic background. The resulting waveforms at $\nu=0.15-0.2$
are strongly reminiscent of the rogue waves on a periodic (i.e.,
elliptic function) background recently discovered in integrable models
such as the nonlinear Schr{\"o}dinger~\cite{peli1} and the modified
Korteweg-de Vries~\cite{peli2} models. While such solutions have
not previously been found, to the best of
our knowledge, in the nonlocal model these results are strongly
suggestive that they exist.
Whether they can be identified in this nonintegrable model in some
closed form remains an outstanding problem for future study. We should
note here that the KM waveform can only be continued up to around
$\nu=0.25$, but not beyond that.

The verification of the accuracy of the numerical solution is given
in Fig.~\ref{fig:Fig11}, in a way similar to what was done before in
Fig.~\ref{fig:Fig12}. Indeed, in this case the error up to $t=5$ does
not grow in all the cases considered beyond O$(10^{-6})$.
The associated growth observed in the ETDRK4 simulations initialized
with the KM initial condition can be attributed to the exponential
instability of the background seeded by the residual numerical error that
eventually will grow to lead to deviations of O$(1)$ at sufficiently
long times (propagation distances).

\subsection{First Doubly Periodic Solution}

\begin{figure}[]
% Fig 5
\center
\subfloat[$\nu=0$]{\includegraphics[width=.24\textwidth]{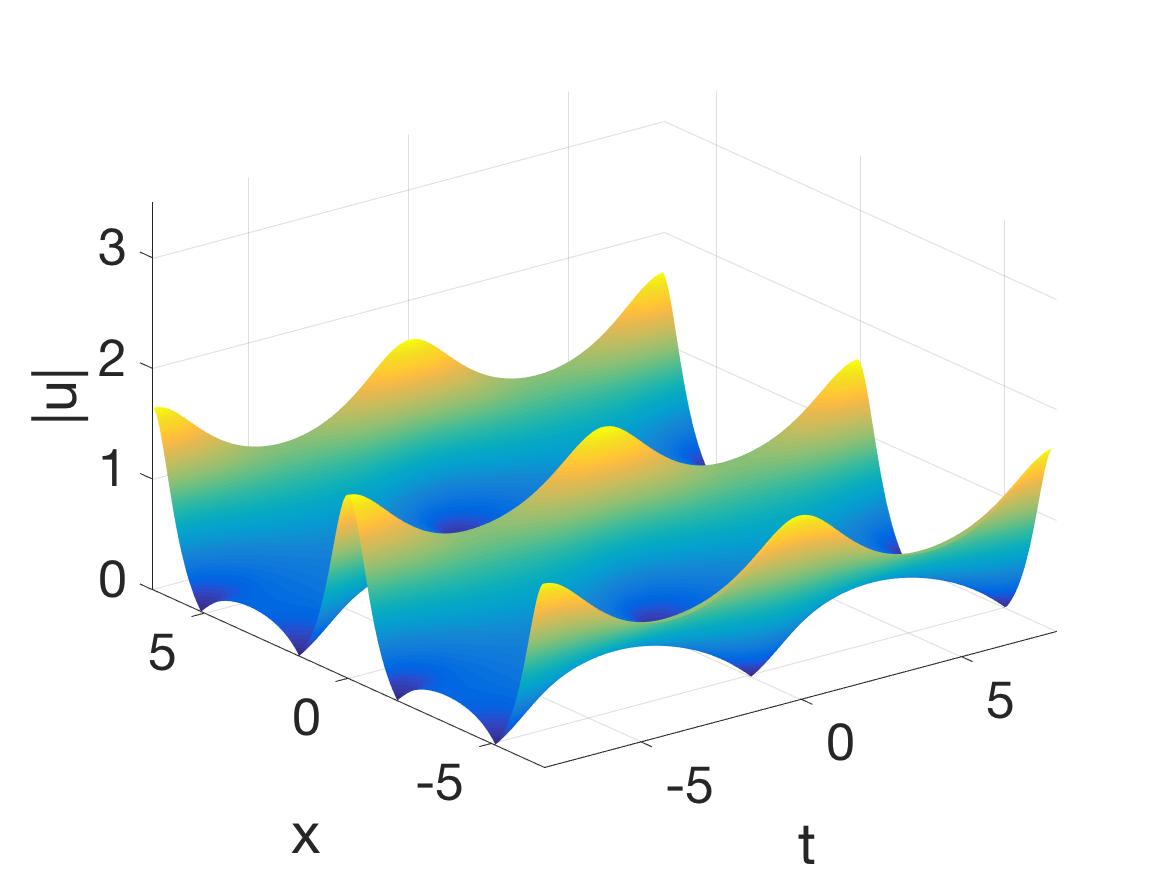}}
\subfloat[$\nu=0.5$]{\includegraphics[width=.24\textwidth]{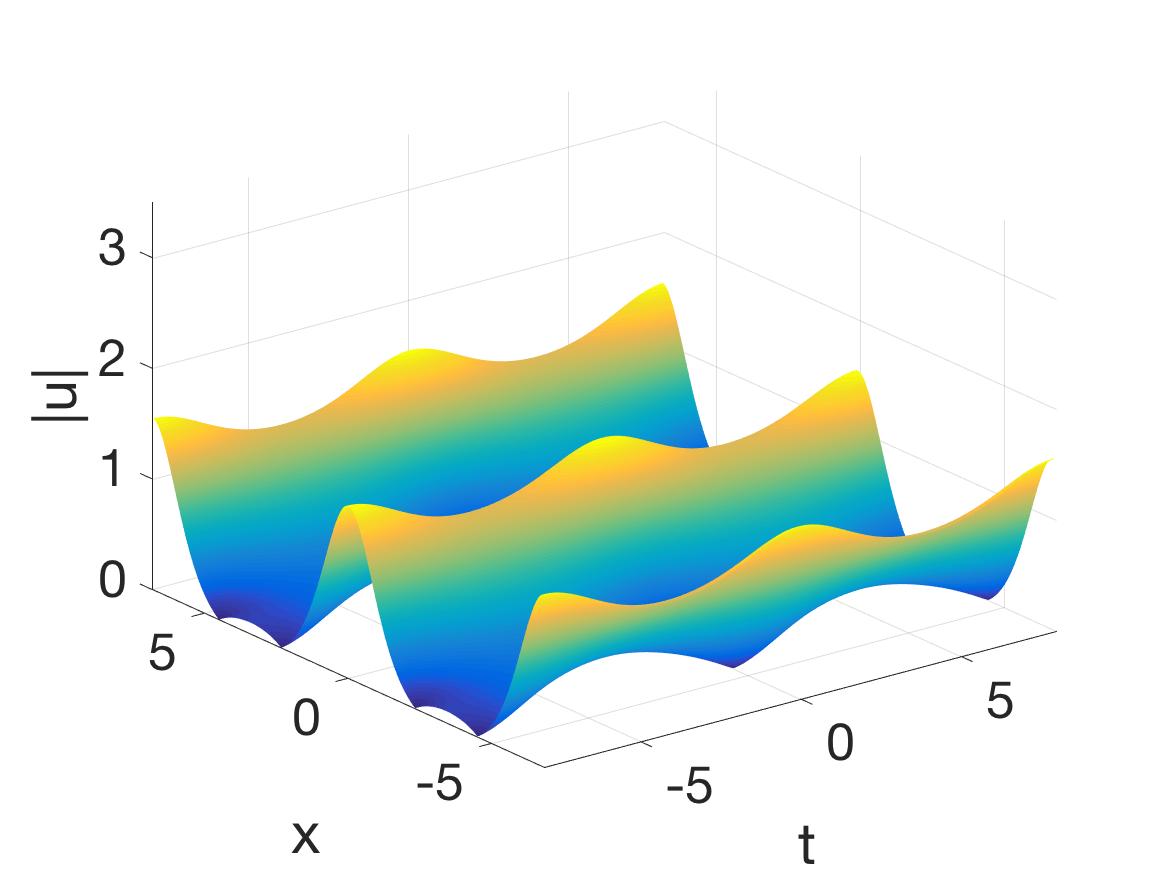}}
\subfloat[$\nu=1$]{\includegraphics[width=.24\textwidth]{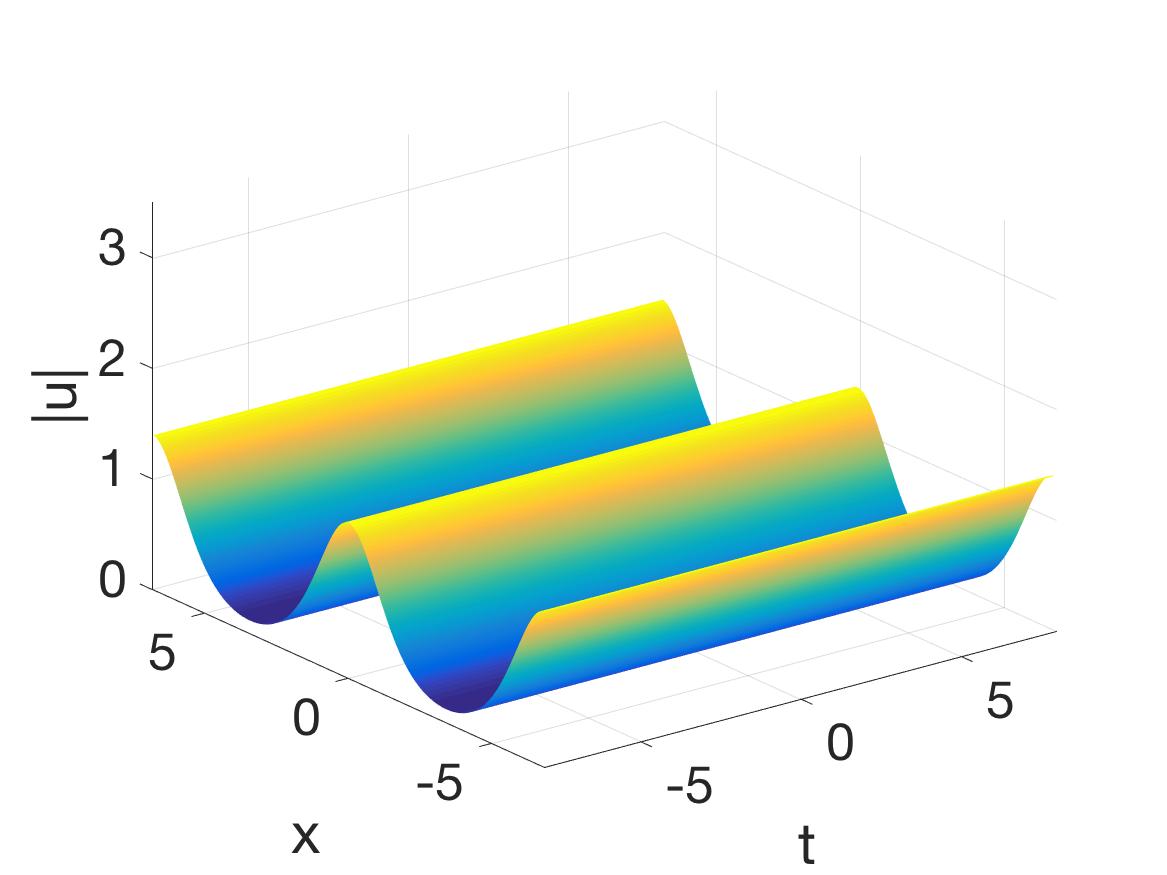}}
\subfloat[$\nu=1.5$]{\includegraphics[width=.24\textwidth]{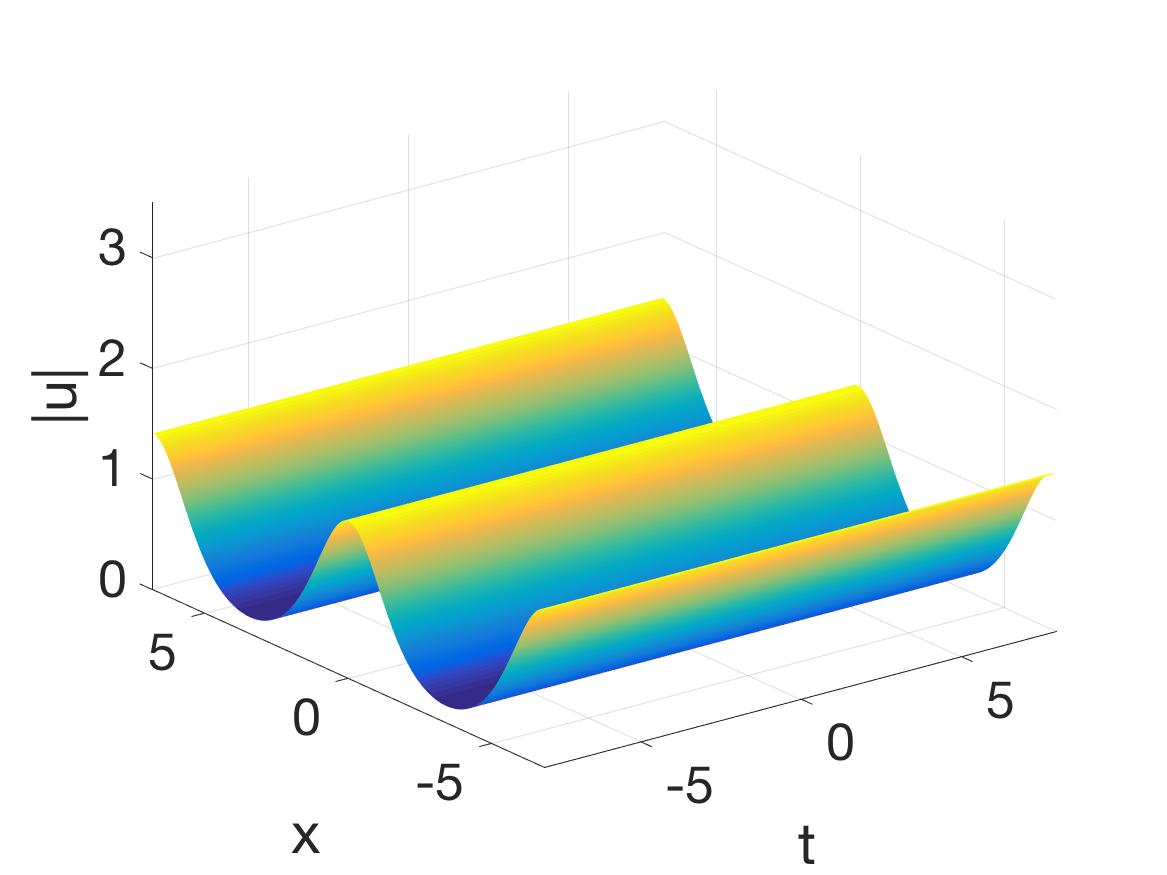}}
%\subfloat[$\nu=2$]{\includegraphics[width=.3\textwidth]{Fig1e.jpg}}

\caption{Continuation over increasing $\nu$ values of the doubly
  periodic first solution of the NLS. It is seen that the doubly periodic
  solution bifurcates into a solution whose (modulus) profile is stationary and
  thus time-independent as $\nu$ is increased.}
\label{fig:Fig1}
\end{figure}

\begin{figure}[]
% Fig 6
\center
\subfloat[$\nu=0$]{\includegraphics[width=.24\textwidth]{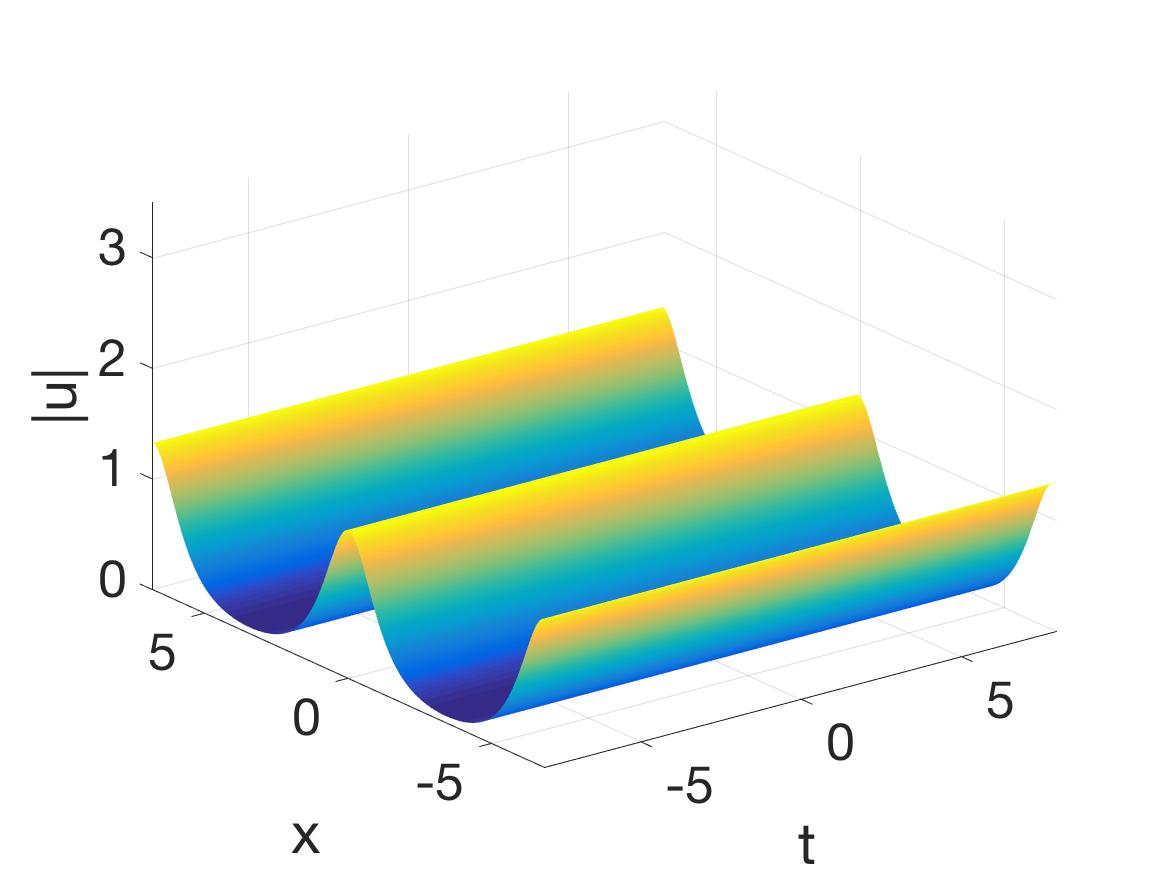}}
\subfloat[$\nu=0.5$]{\includegraphics[width=.24\textwidth]{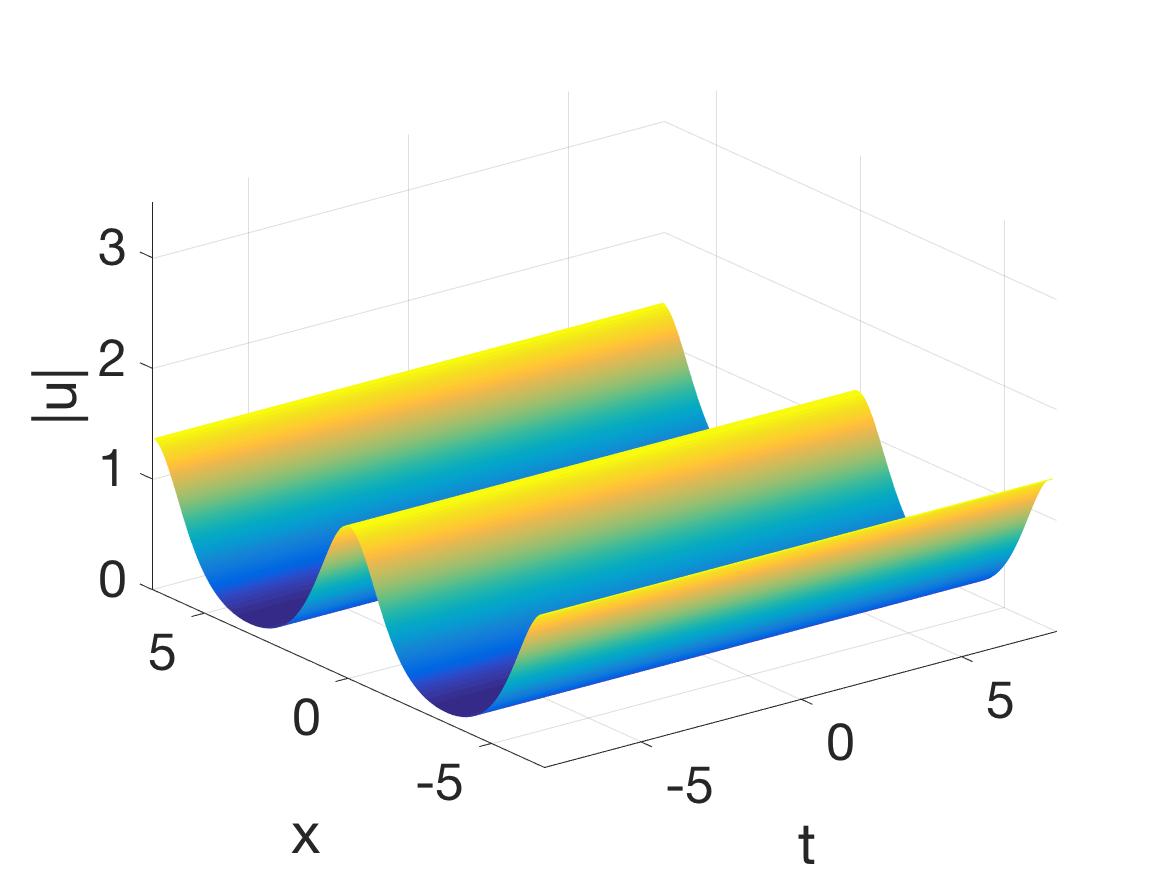}}
\subfloat[$\nu=1$]{\includegraphics[width=.24\textwidth]{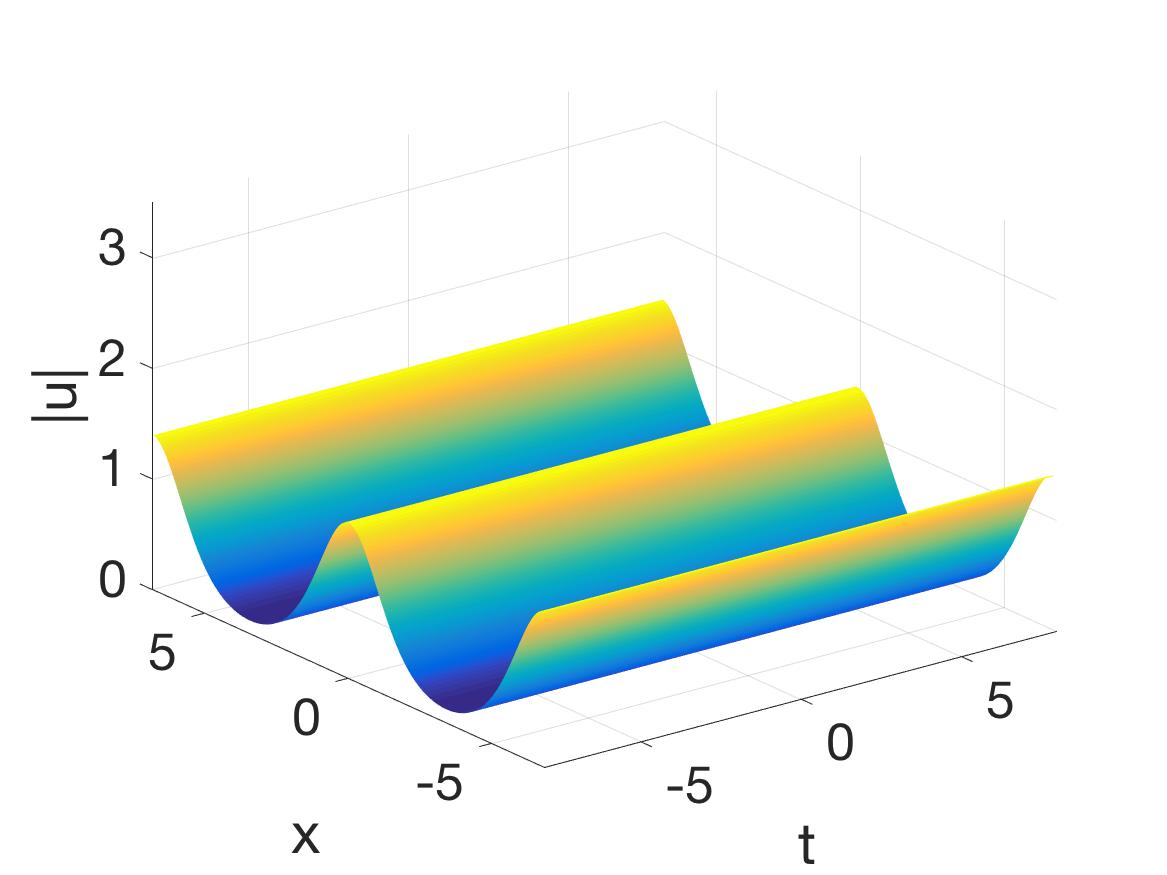}}
%\subfloat[$\nu=1.5$]{\includegraphics[width=.24\textwidth]{Fig2d.jpg}}
\subfloat[$\nu=2$]{\includegraphics[width=.24\textwidth]{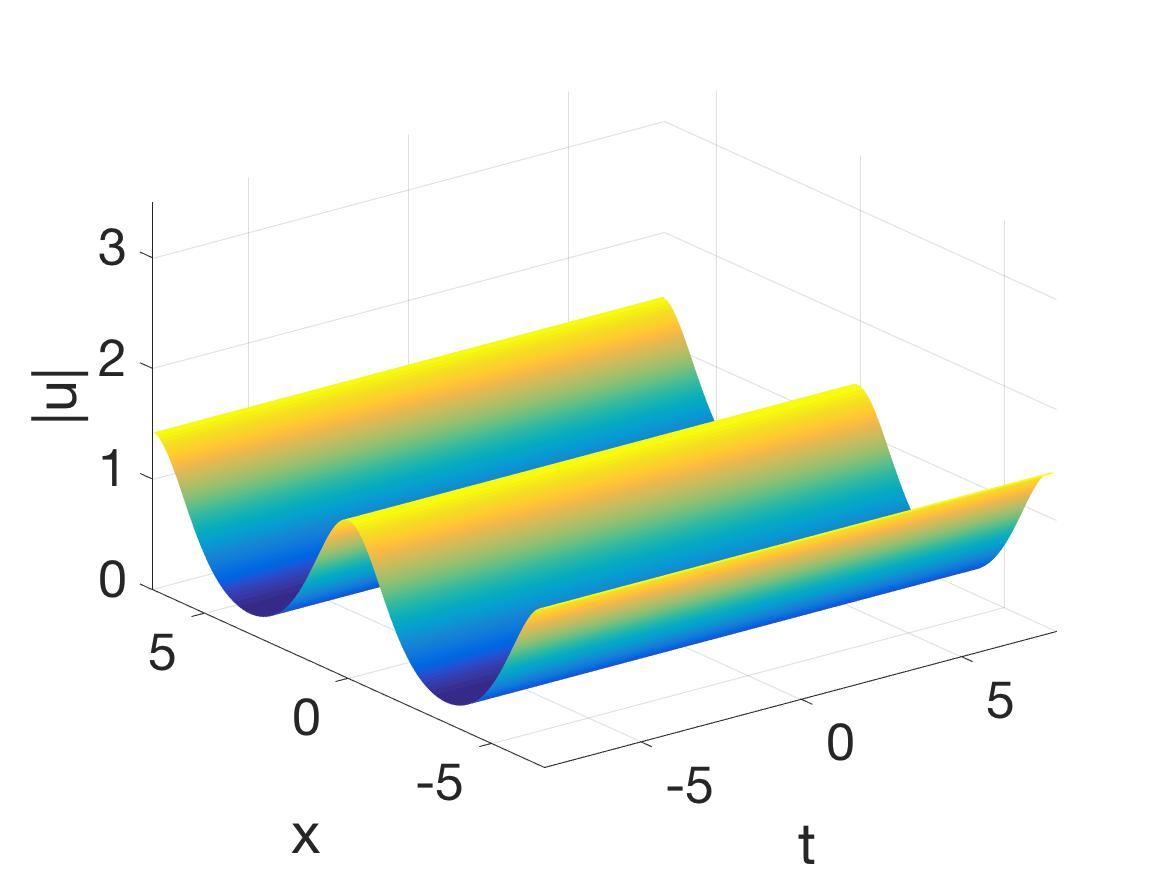}}

\caption{Convergence to the stationary, periodic in space solution, obtained upon
using as initial condition the solution in Fig.~\ref{fig:Fig1}(e), and decrease of
$\nu$. This periodic solution apparently exists all the way to the local NLS limit.}
\label{fig:Fig2}
\end{figure}

\begin{figure}[]
% Fig 7
\center
\subfloat[]{\includegraphics[width=.3\textwidth]{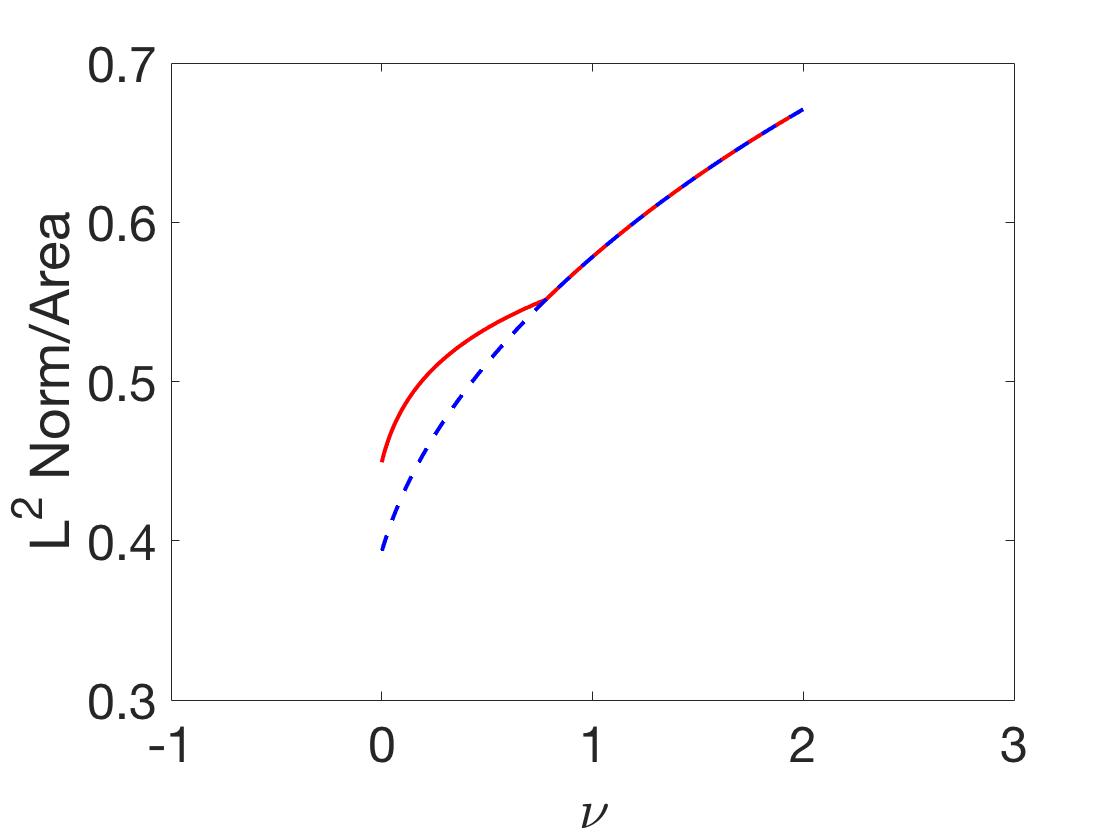}}

\caption{Bifurcation diagram showing the doubly periodic solution
  (solid) and the constant profile solution (dashed) as $\nu$
  varies. Clearly, as $\nu$ increases the two branches merge with each
other, i.e., the periodic (in time) orbit emerges from the stationary solution.}
\label{fig:Fig3}
\end{figure}

We now turn to the doubly periodic solution Eq.~(\ref{neq5}) for
which the results are presented in Figs.~\ref{fig:Fig1}-\ref{fig:Fig9};
here we have chosen $\kappa=0.8$. As can be seen in Fig.~\ref{fig:Fig1}, the amplitude
of the solution is initially doubly periodic in both space and time but gradually,
as the nonlocality parameter is increased, becomes singly periodic in
space. By this we mean that the modulus of the solution becomes
time-independent (i.e., we exclude phase factors that can be
eliminated by means of a gauge transformation).
%That being said, the solution itself is not time-independent, having a phase factor of the form $\exp( \frac{2 \pi i t}{T})$, where $T$ is the time-period of the original doubly periodic solution. This phase factor can be factored out however, making the time-independent amplitude a solution of Eqs.~\eqref{neq1}, \eqref{neq2}, with $\mu = \frac{1}{2}+ \frac{2 \pi}{T}$, in it's own right.

This naturally suggests the question of whether this stationary,
periodic in space solution of the nonlocal problem emerges only in the
nonlocal
model or exists in the local limit. Indeed, continuing the stationary
solution ``downward'' (i.e., for decreasing $\nu$) one can see that
it bifurcates from a stationary (elliptic function) solution at the
local
NLS limit, given explicitly by
\begin{equation}
u = \exp (i\frac{\alpha t}{T}) \; A \; \text{dn}\big( Ax, \sqrt{2 + \frac{2 \alpha-1}{A^2}}\big)
\end{equation}
where $\alpha = \frac{2 \pi}{T}$ and $A \approx 1.43$.
 The case examples of this state (for the same values of
$\nu$
as in Fig.~\ref{fig:Fig1}) are shown in Fig.~\ref{fig:Fig2}. Indeed,
the relevant bifurcation diagram illustrating the merger of the
space-time periodic solution with the space-periodic one is shown
in Fig.~\ref{fig:Fig3}. The emergence of a solution with a finite
periodicity
from a stationary one suggests a Hopf scenario in the Hamiltonian
system at hand.

Lastly, here too, we examine the growth of the residual and find it to
be very small ($(10^{-8})$ at the highest), when considering this class of solutions
for different $\nu$'s even up to $\nu=2$ (for which the solution is stationary).
The relevant results are presented in Fig.~\ref{fig:Fig9}
for the periodic solution in time (that turns stationary as $\nu$ is
increased). We show prototypical examples for the space-time periodic
solution in panels (a)-(b), then for higher values of $\nu$ where the
waveform acquires a stationary modulus in (c)-(d). Also for
completeness,
in panels (e)-(f) we go back down to lower values of $\nu$ along
the branch of solutions of stationary modulus (i.e., periodic only
in space) and examine the error in this case as well, confirming that
it remains quite small during the evolution interval considered. 
%and in
%Fig.~\ref{fig:Fig10} for the one whose modulus is time-independent.

\begin{figure}[]
% Fig 8
\center
\subfloat[$\nu=0$]{\includegraphics[width=.3\textwidth]{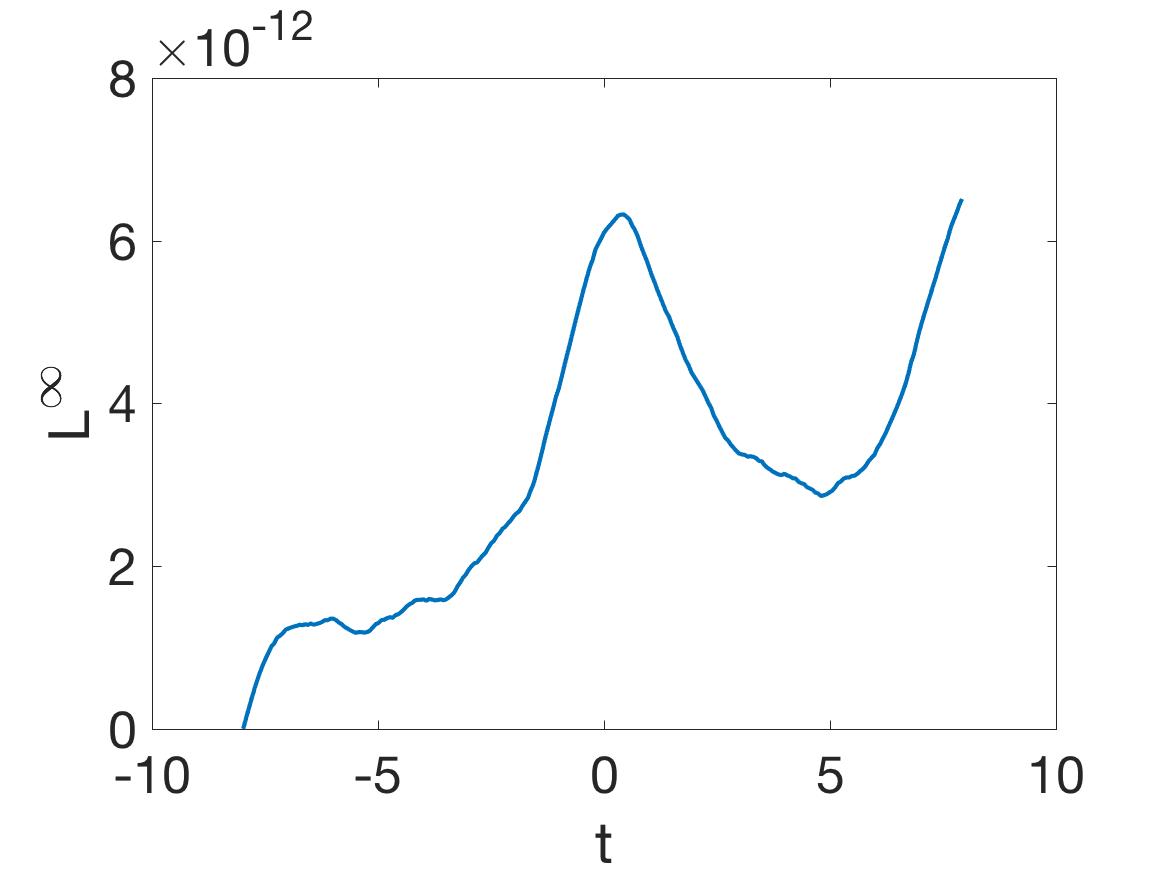}}
\subfloat[$\nu=0.5$]{\includegraphics[width=.3\textwidth]{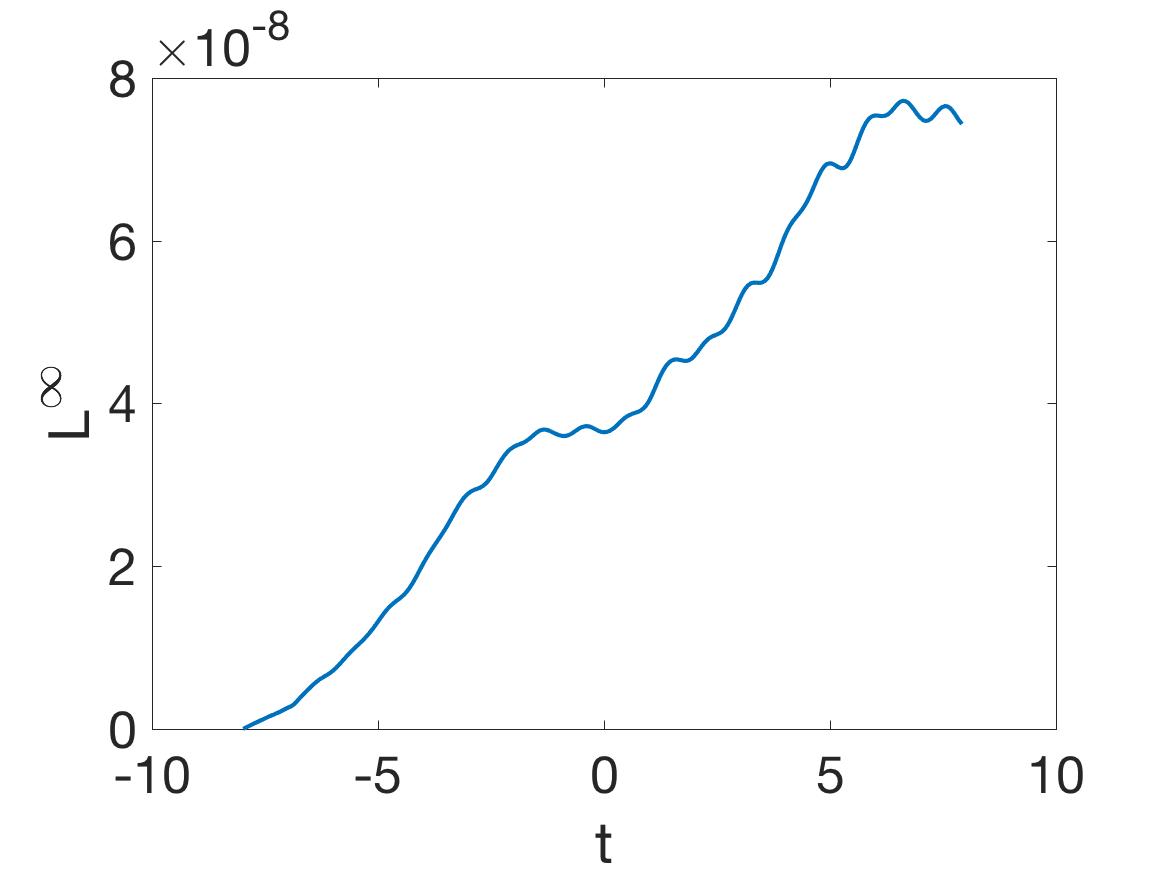}}
\subfloat[$\nu=1$]{\includegraphics[width=.3\textwidth]{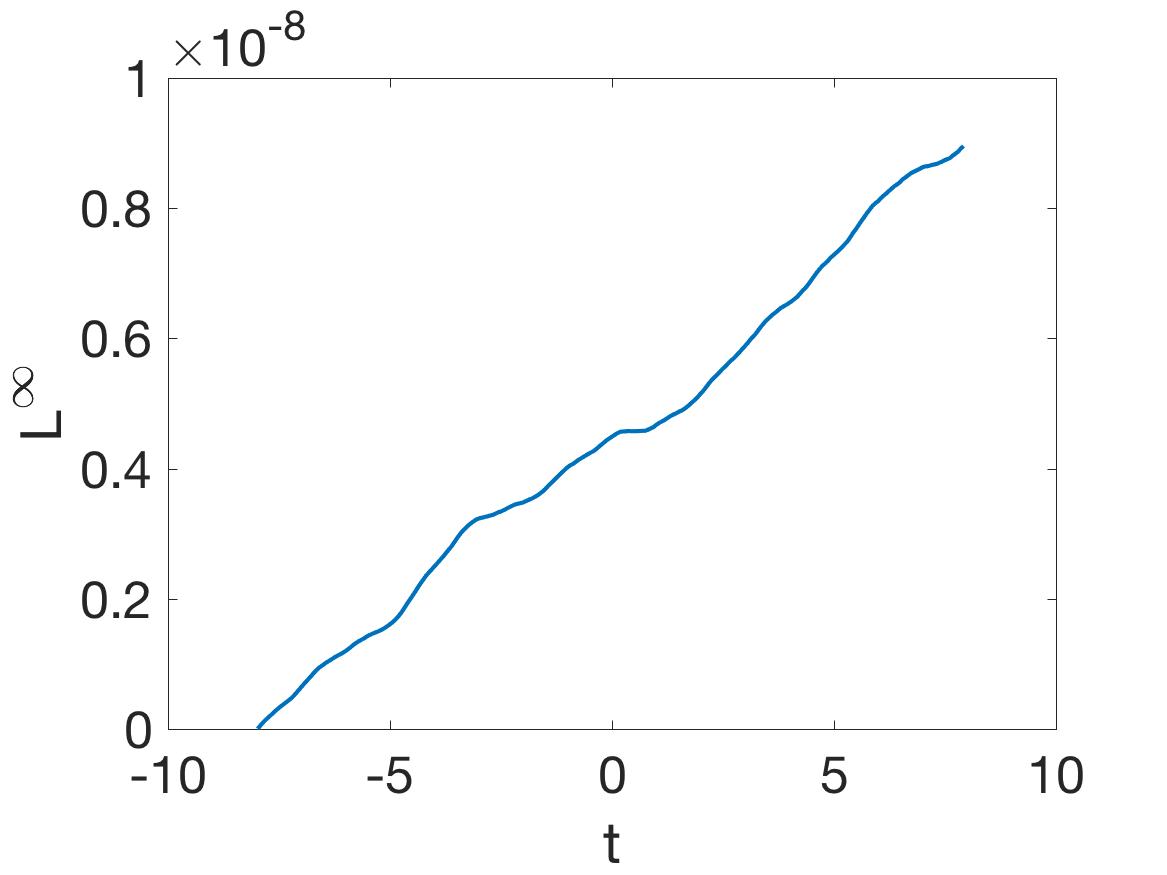}}
\\
%\subfloat[$\nu=1.5$]{\includegraphics[width=.3\textwidth]{Fig9d.jpg}}
\subfloat[$\nu=2$]{\includegraphics[width=.3\textwidth]{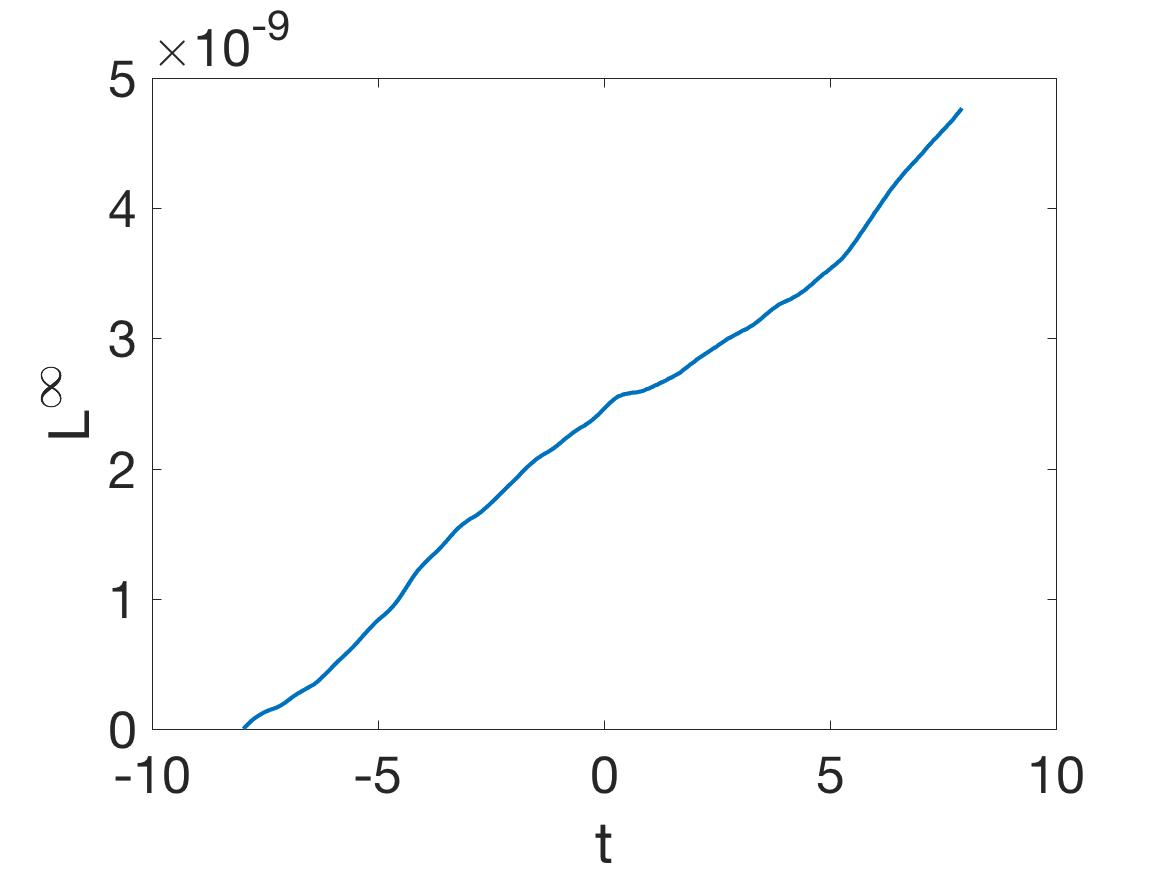}}
\subfloat[$\nu=0.5$]{\includegraphics[width=.3\textwidth]{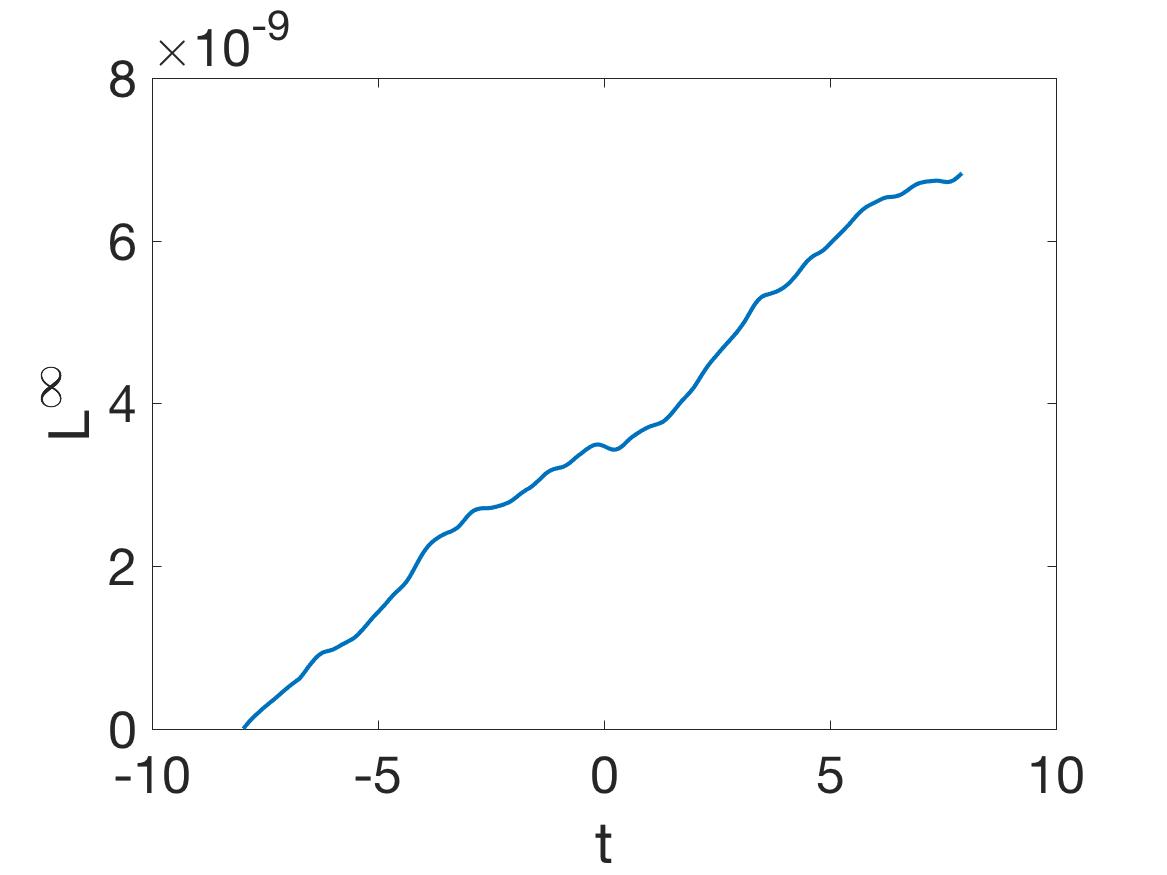}}
\subfloat[$\nu=0$]{\includegraphics[width=.3\textwidth]{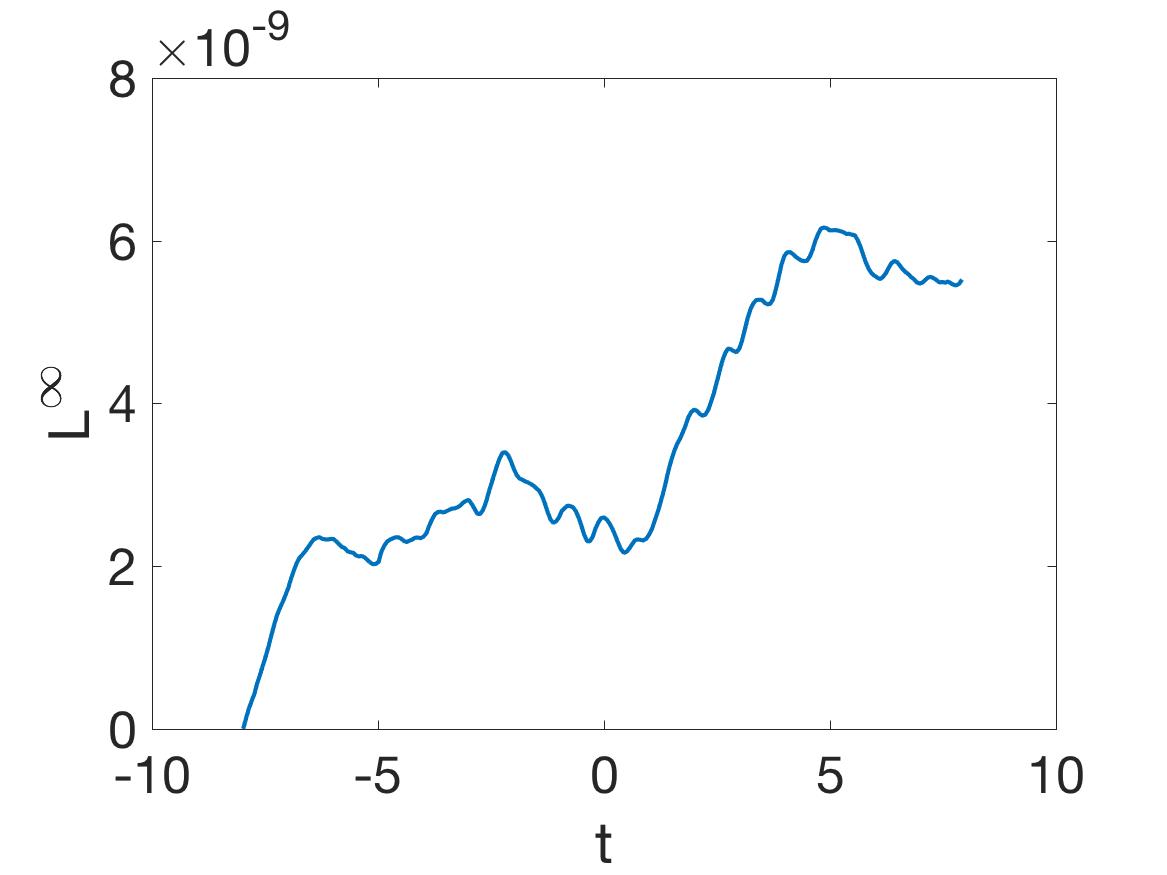}}
\caption{$L^\infty$ error between the solution obtained by the
  Newton-CG method (as shown in Fig.~\ref{fig:Fig1}) vs the solution
  obtained via the ETDRK4  method. Different values of $\nu$ up to
  $\nu=2$ are shown. Recall that in panels (c)-(f) the solution is, in
fact, stationary. In panels (e)-(f) the error is shown for the
solutions continued
back down to lower values of $\nu$ along the branch of stationary
(periodic in space) modulus.}
\label{fig:Fig9}
\end{figure}

%\begin{figure}[]
% Fig 9
%\center
%\subfloat[$\nu=0$]{\includegraphics[width=.24\textwidth]{Fig10a.jpg}}
%\subfloat[$\nu=0.5$]{\includegraphics[width=.24\textwidth]{Fig10b.jpg}}
%\subfloat[$\nu=1$]{\includegraphics[width=.24\textwidth]{Fig10c.jpg}}
%%\subfloat[$\nu=1.5$]{\includegraphics[width=.3\textwidth]{Fig10d.jpg}}
%\subfloat[$\nu=2$]{\includegraphics[width=.24\textwidth]{Fig10e.jpg}}
%\caption{$L^\infty$ error between the solution obtained by the
%  Newton-CG method (shown in Fig.~\ref{fig:Fig2}) vs the solution
%  obtained via the ETDRK4  method. These results are for the genuinely
%stationary solution from which the one which is periodic in time
%emerges. }
%\label{fig:Fig10}
%\end{figure}

\subsection{Second Doubly Periodic Solution}

In a similar vein, we now examine the continuation of the local
solution Eq.~(\ref{neq6}) to the nonlocal regime. We find a similar
phenomenology as a result of the continuation as in the previous section.
Namely, the doubly periodic solution, as $\nu$ is increased, turns to a singly
periodic one in the case of sufficiently large $\nu$; see Fig.~\ref{fig:Fig4}.
On the other hand, starting at $\nu=2$ and
continuing the stationary (modulus) solution down to
$\nu=0$, we find that the relevant waveform exists for all $\nu$
down to the local limit, as illustrated in Fig.~\ref{fig:Fig5}.%
That is to say, there is again a bifurcation diagram
Fig.~\ref{fig:Fig6} illustrating the Hopf type emergence of the periodic
(in time) orbit from the stationary one (for decreasing $\nu$).

An additional interesting observation is that this stationary solution in
the NLS limit ($\nu = 0$) degenerates to an explicit cnoidal solution in the form of
\begin{eqnarray}
  u(x)=
  A \; \text{cn} \left(\sqrt{-1+2A^2} \; x, ~\sqrt{\frac{2A^2}{2(-1+2A^2)}}\right),
  \label{neq10}
  \end{eqnarray}
with $A\approx 1.62$.
Further, we can also arrive at an exact expression for the stationary state as $\nu \to \infty$. Inspired by the numerics, we can judiciously guess a solution of the form
\[\begin{cases}
u = A \cos(Bx), \\
\theta = C \cos(Dx) + E,
\end{cases}
\]
and use it in Eq.~(\ref{neq2}). Doing so, and matching terms, one gets conditions on the constants, namely
\[B=\frac{D}{2}, \quad E=\frac{A^2}{2}, \quad C= \frac{\beta}{2}\frac{A^2}{D^2+\beta},\]
where $\beta = \frac{2}{\nu}$. Equivalently,
\[\begin{cases}
u = A \cos(\frac{D}{2}x) \\
\theta = \frac{\beta}{2}\frac{A^2}{D^2+\beta} \cos(Dx) + \frac{A^2}{2}.
\end{cases}
\]
Plugging this into the LHS of Eq.~(\ref{neq1}) and simplifying yields
\begin{align}
&i\frac{\partial u}{\partial t} + \frac{1}{2} \frac{\partial^2 u}{\partial x^2} + \theta u - \frac{1}{2} u  \nonumber \\
 &=-\Big(\frac{D^2}{8} + \frac{1}{2} - \frac{A^2}{2}\Big) A \cos \left(\frac{D}{2}x\right)
 + \Big(\frac{\beta A^2}{2} \frac{A}{D^2 +
 \beta} \Big)\cos (Dx)\cos \left(\frac{D}{2}x\right).
\end{align}
Taking $\beta \to 0$, or equivalently $\nu \to \infty$, the above reduces to
%\[
  $-\big(\frac{D^2}{8} + \frac{1}{2} - \frac{A^2}{2}\big) A \cos
  \left(\frac{D}{2}x\right)$
%\]
which we can make zero provided we choose
\[\frac{D^2}{8} + \frac{1}{2} - \frac{A^2}{2}=0.\]

This has two consequences. The first is that, as $\nu \to \infty$, we have the exact solution:
\[\begin{cases}
u = A \cos(\sqrt{A^2-1} x) \\
\theta =\frac{A^2}{2}
\end{cases}
\]
The second, related one, is that under the limit $\nu \to \infty$ the nonlocal NLS reduces
to the following \emph{linear} Schr\"odinger equation: %given by
\[\frac{1}{2} \frac{\partial^2 u}{\partial x^2} + \frac{A^2}{2}u - \frac{1}{2} u =0.\]
From the bifurcation diagram  in Fig.~\ref{fig:Fig6}, as $\nu$
increases, we have found
that the amplitude of $u$ approaches the constant value of $A \approx 1.92$.

It is relevant to remark here that the numerically obtained periodic
in space solution is well approximated by the above asymptotic
functional form. At $\nu=6$ the pointwise error between the two is approximately
$2 \times 10^{-2}$.
In fact, our numerical observations suggest that one can approximate
the numerically obtained
periodic in space solution by the functional form:
\[u=A \text{cn}(Bx, C)\]
for suitably chosen $A,B,C$. In fact, with this approximation, the
pointwise error can be made between $10^{-3}$ and $10^{-6}$ for all $\nu$ in the interval
$0 \leq \nu \leq 6$. This suggests the possibility of seeking suitable
elliptic function solutions as potentially exact waveforms of the
nonlocal model. This merits a separate investigation beyond
the confines of the present work. Lastly, we note that for both
branches
of solutions, we present the diagnostic of the $L^{\infty}$ norm of
the
deviation from the numerically identified solution when propagating
the corresponding initial condition with ETDRK4 in
Fig.~\ref{fig:Fig13}. This is similar to Fig.~\ref{fig:Fig9} with
the top panels pertaining to the space-time periodic branch, while
the bottom ones arise for the stationary modulus, periodic in space
branch present for large
$\nu$,
but which can be continued down to lower values of $\nu$.
%-\ref{fig:Fig14}.
Once again the relevant residual grows due to the instability of the
background, yet remains bounded by $10^{-8}$ over the time scales
considered.

\begin{figure}[]
% Fig 10
\center
\subfloat[$\nu=0$]{\includegraphics[width=.24\textwidth]{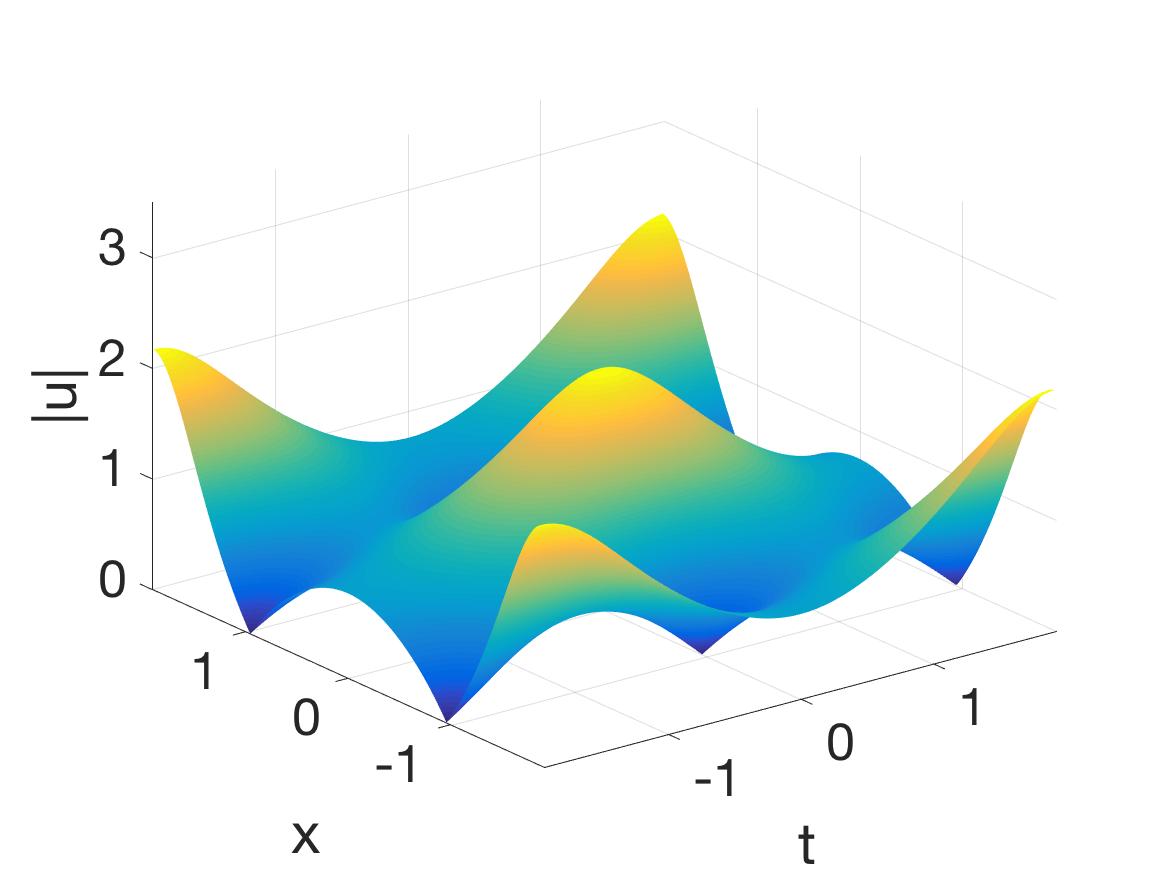}}
\subfloat[$\nu=1.5$]{\includegraphics[width=.24\textwidth]{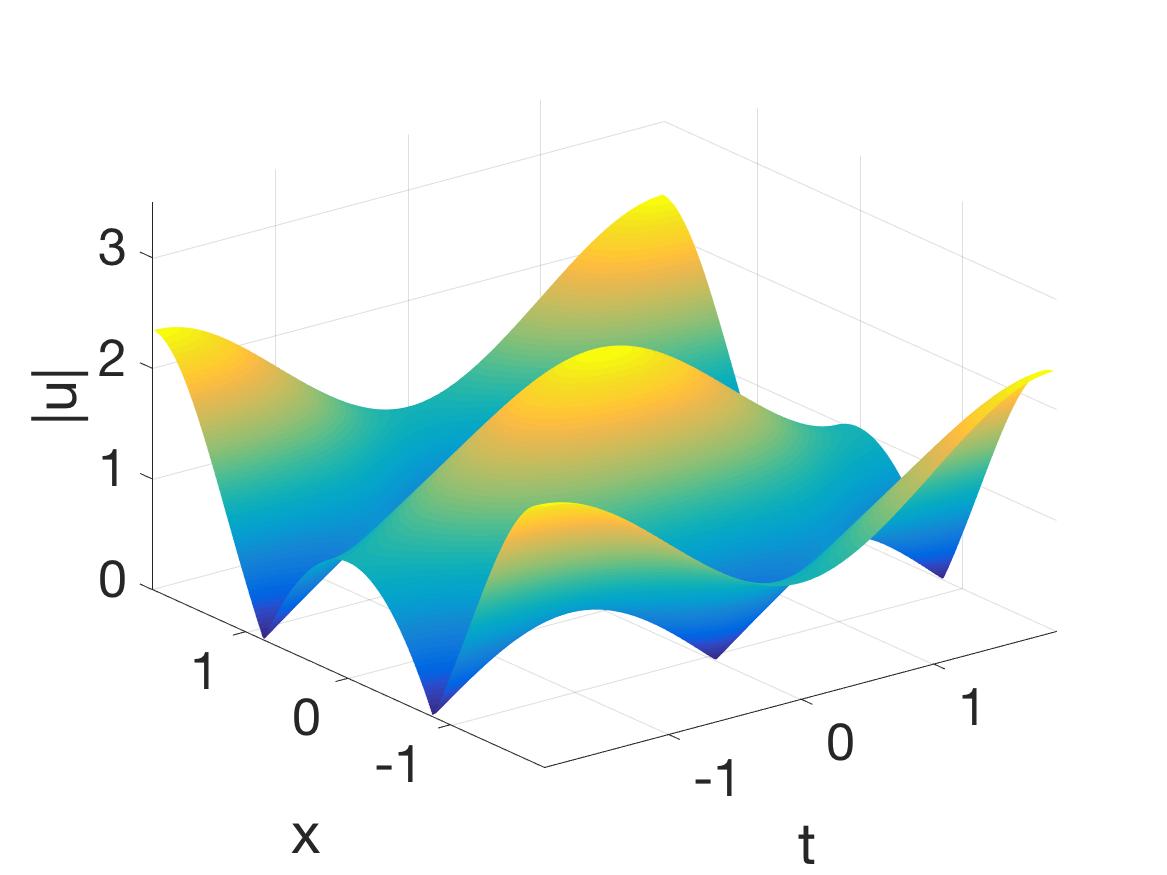}}
\subfloat[$\nu=3$]{\includegraphics[width=.24\textwidth]{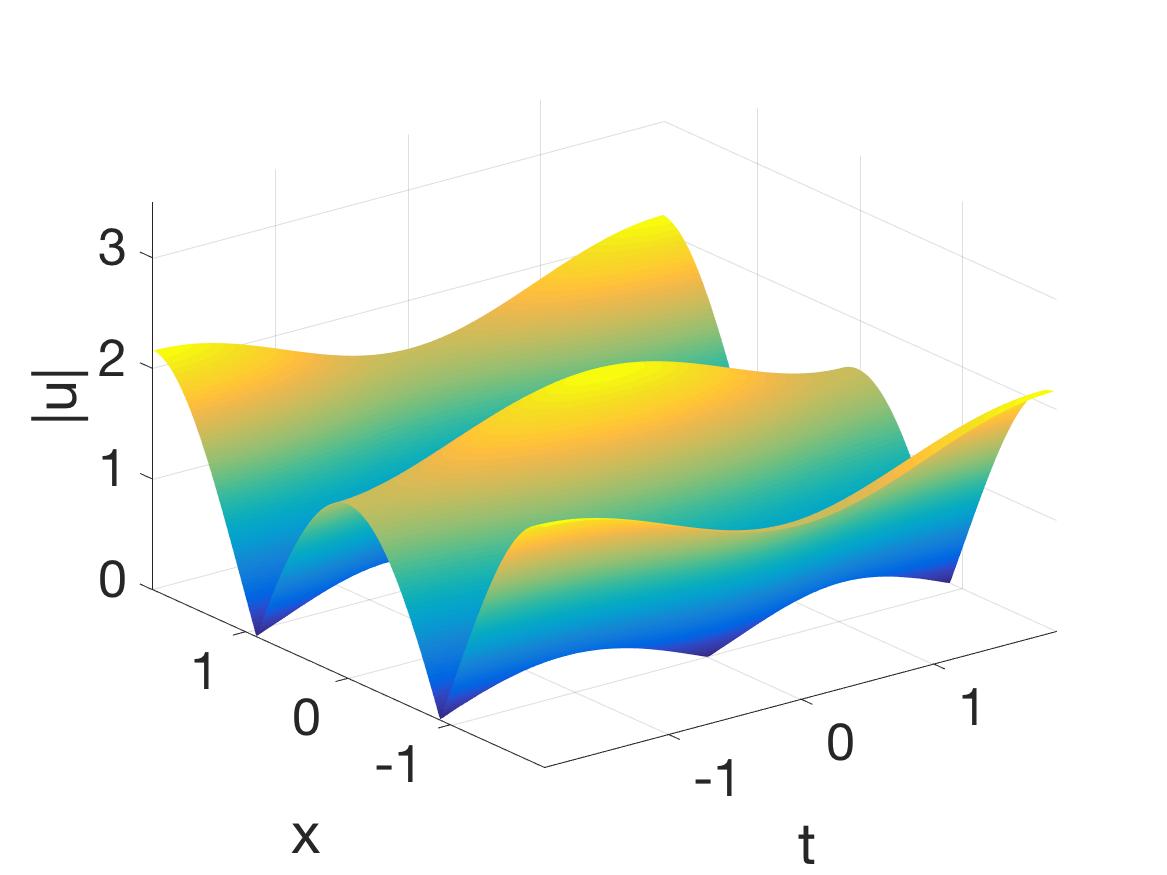}}
%\subfloat[$\nu=4.5$]{\includegraphics[width=.24\textwidth]{Fig7d.jpg}}
\subfloat[$\nu=6$]{\includegraphics[width=.3\textwidth]{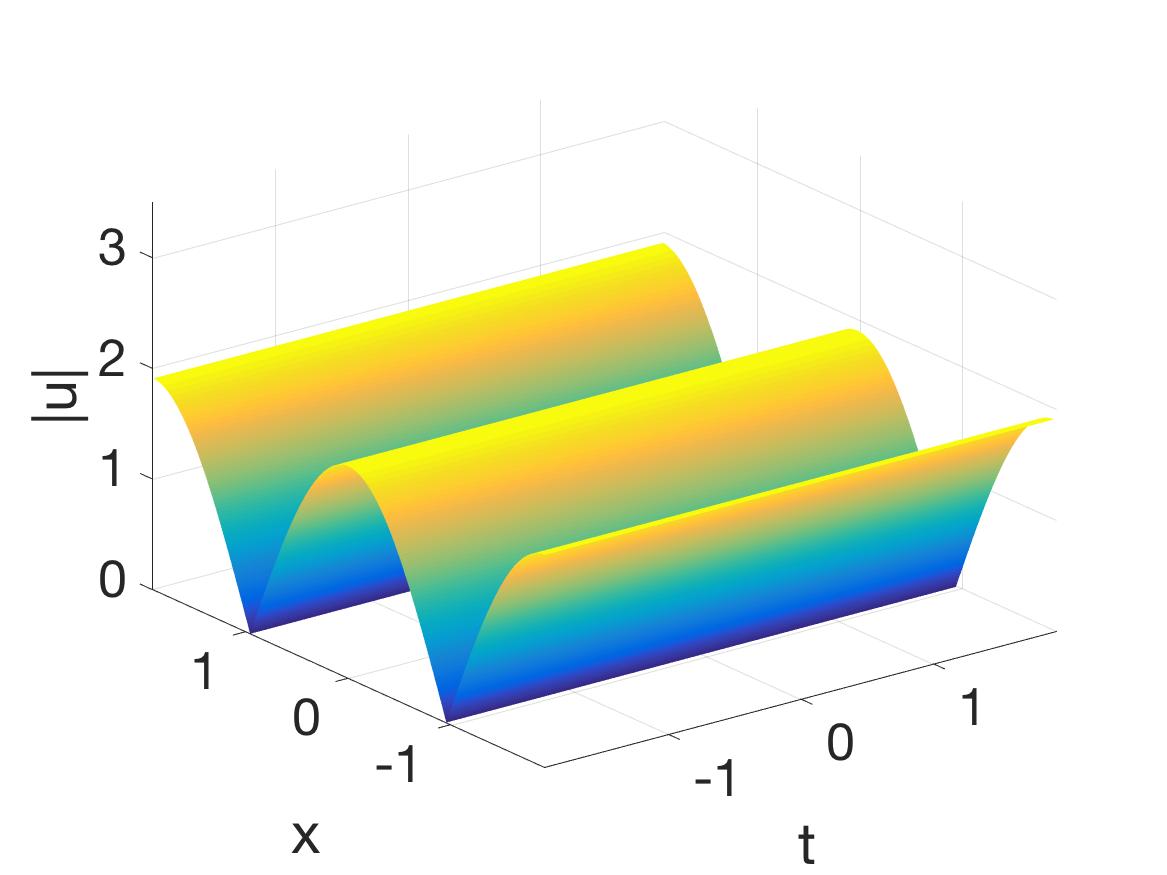}}
\caption{Continuation over increasing $\nu$ values of the second
  doubly periodic solution of the NLS. It is seen that the doubly
  periodic solution bifurcates into a solution whose profile is
  constant in time
as $\nu$ is increased.}
\label{fig:Fig4}
\end{figure}

\begin{figure}[]
% Fig 11
\center
\subfloat[$\nu=0$]{\includegraphics[width=.24\textwidth]{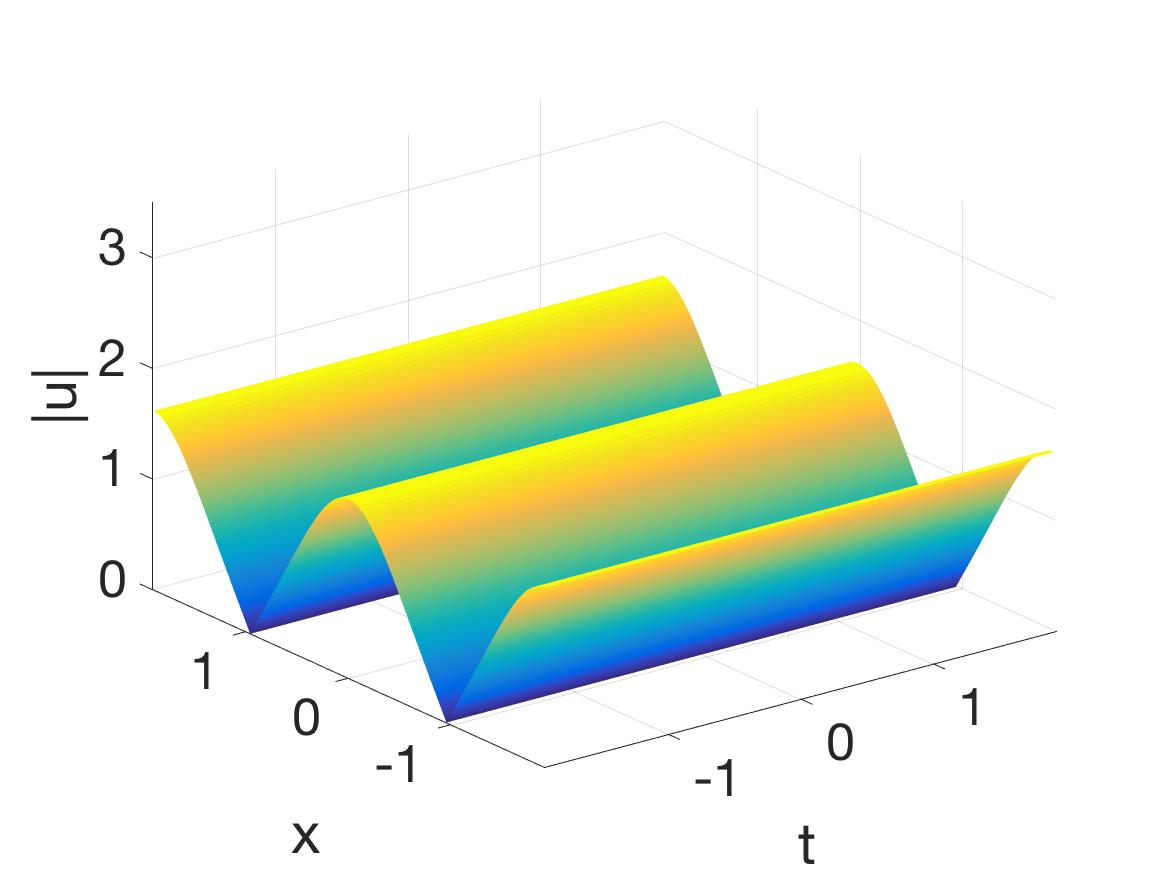}}
\subfloat[$\nu=1.5$]{\includegraphics[width=.24\textwidth]{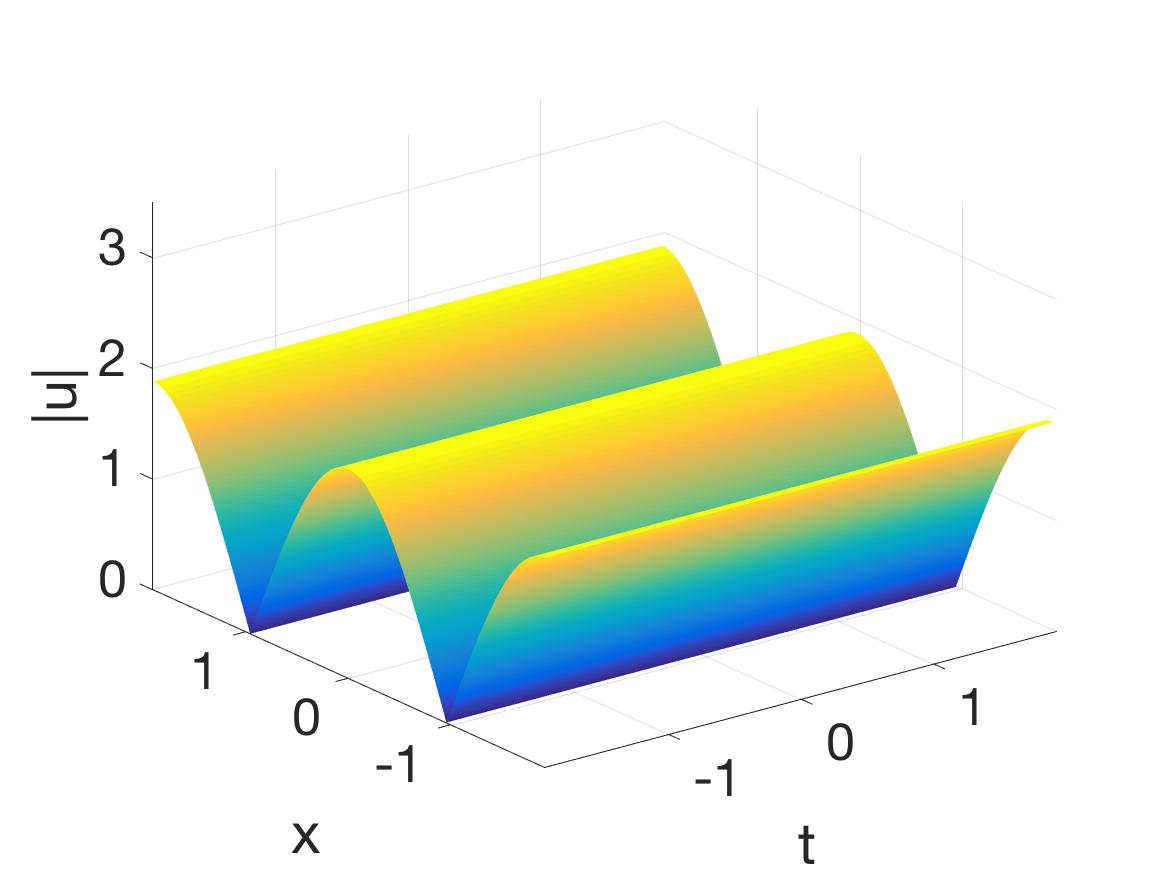}}
\subfloat[$\nu=3$]{\includegraphics[width=.24\textwidth]{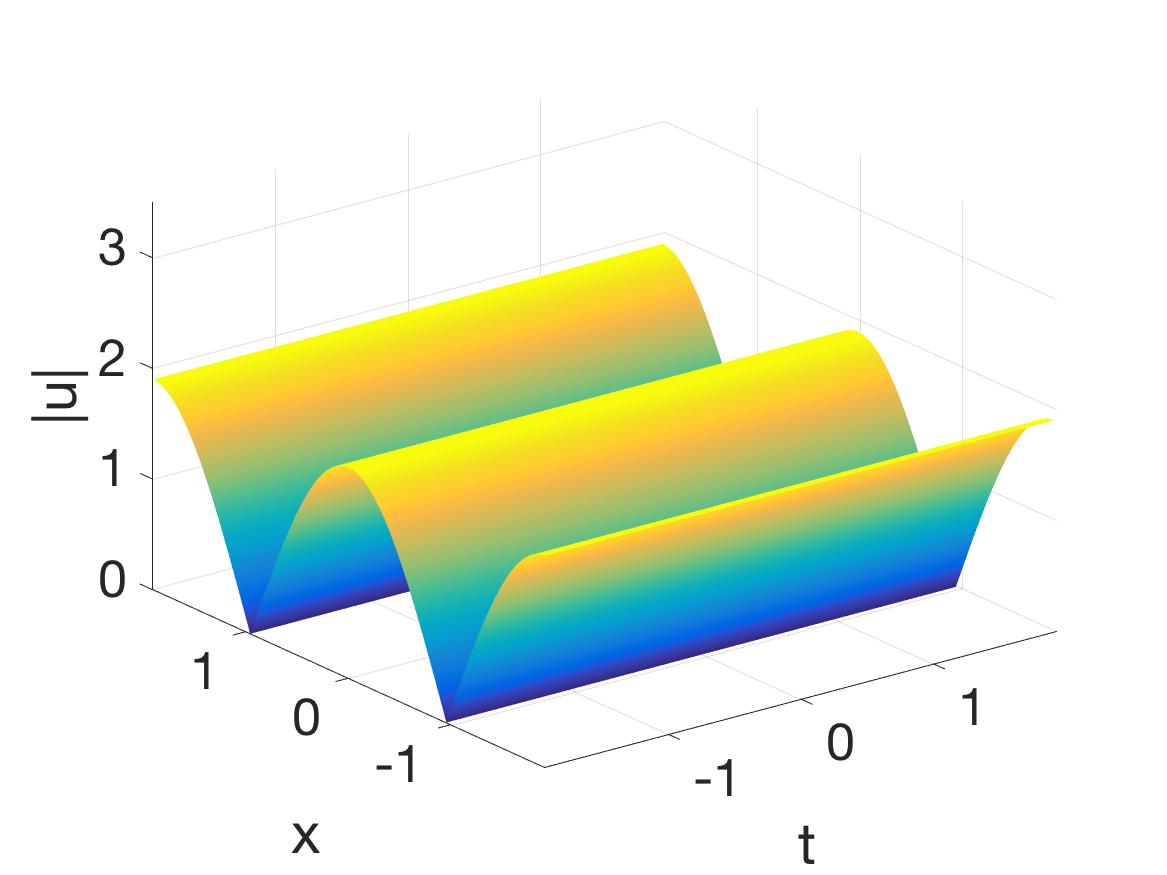}}
%\subfloat[$\nu=4.5$]{\includegraphics[width=.3\textwidth]{Fig8d.jpg}}
\subfloat[$\nu=6$]{\includegraphics[width=.24\textwidth]{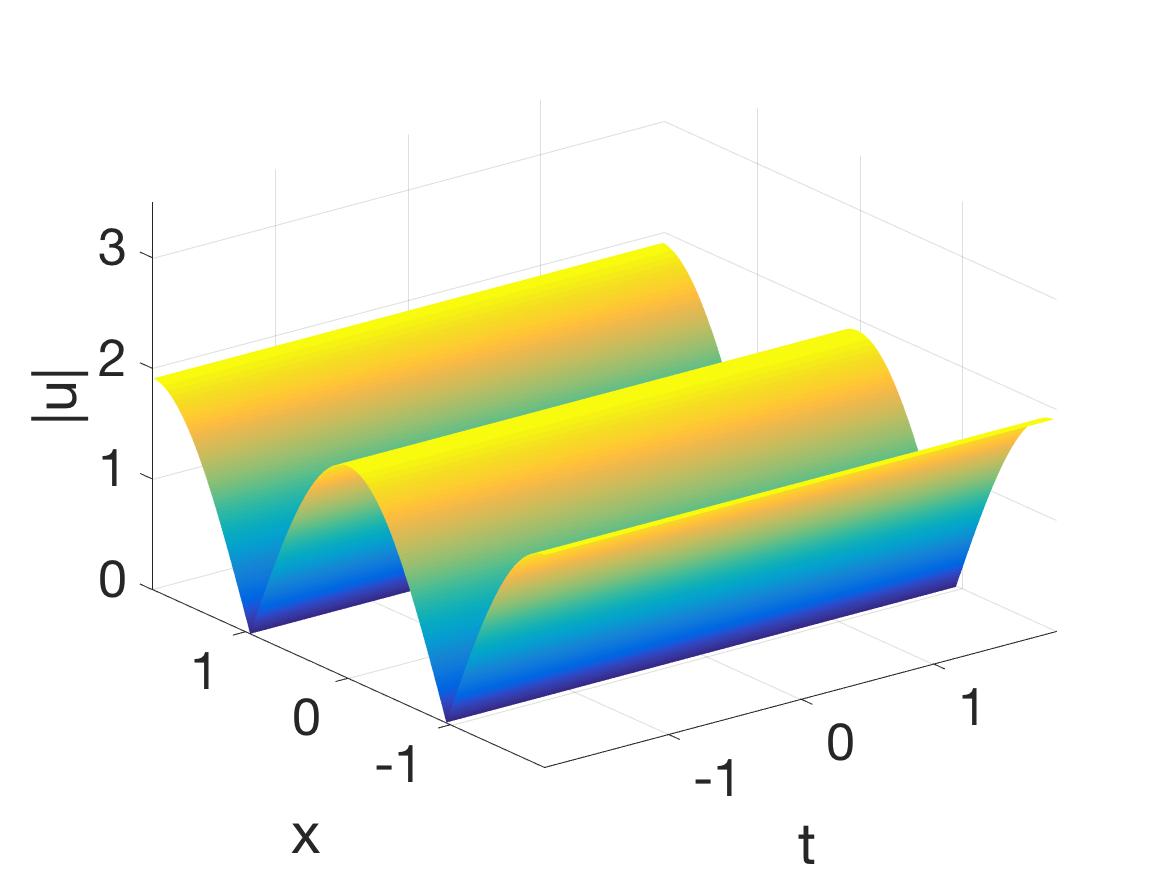}}

\caption{Using as initial condition the solution in
  Fig.~\ref{fig:Fig4}(e), we decrease $\nu$ and
  find the continuation of the stationary branch down to the NLS limit
  of $\nu=0$.}
\label{fig:Fig5}
\end{figure}

\begin{figure}[]
% Fig 12
\center
\subfloat[]{\includegraphics[width=.3\textwidth]{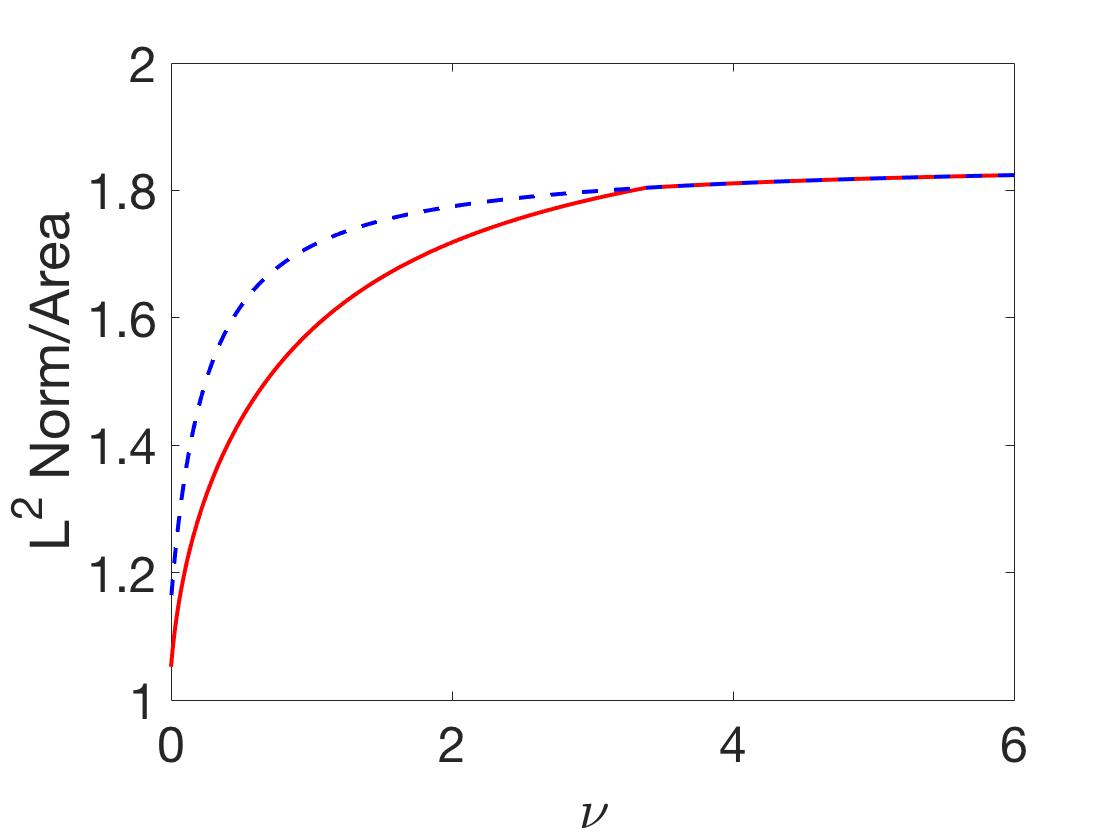}}

\caption{Bifurcation diagram showing the second doubly periodic solution (solid) and the constant profile solution (dashed)
as $\nu$ varies.}
\label{fig:Fig6}
\end{figure}

\begin{figure}[]
%Fig 13
\center
\subfloat[$\nu=0$]{\includegraphics[width=.3\textwidth]{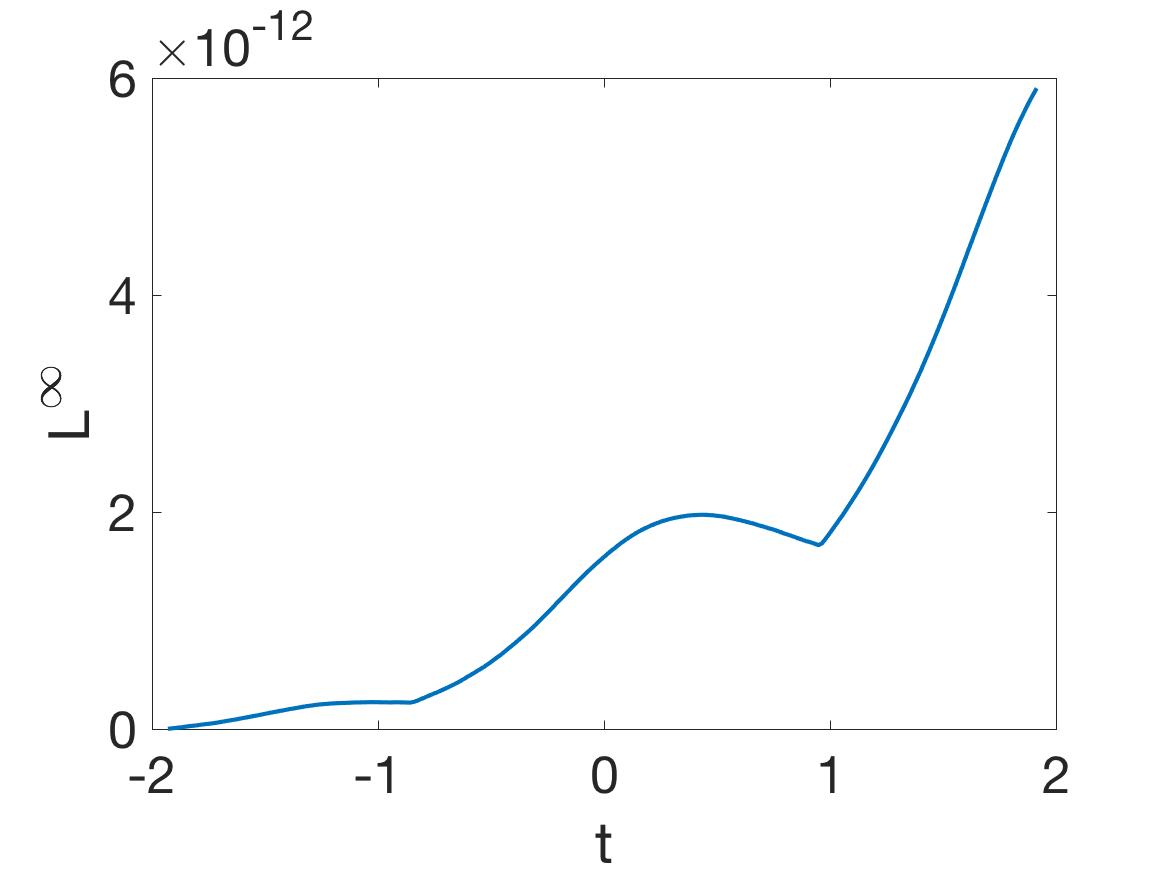}}
\subfloat[$\nu=1.5$]{\includegraphics[width=.3\textwidth]{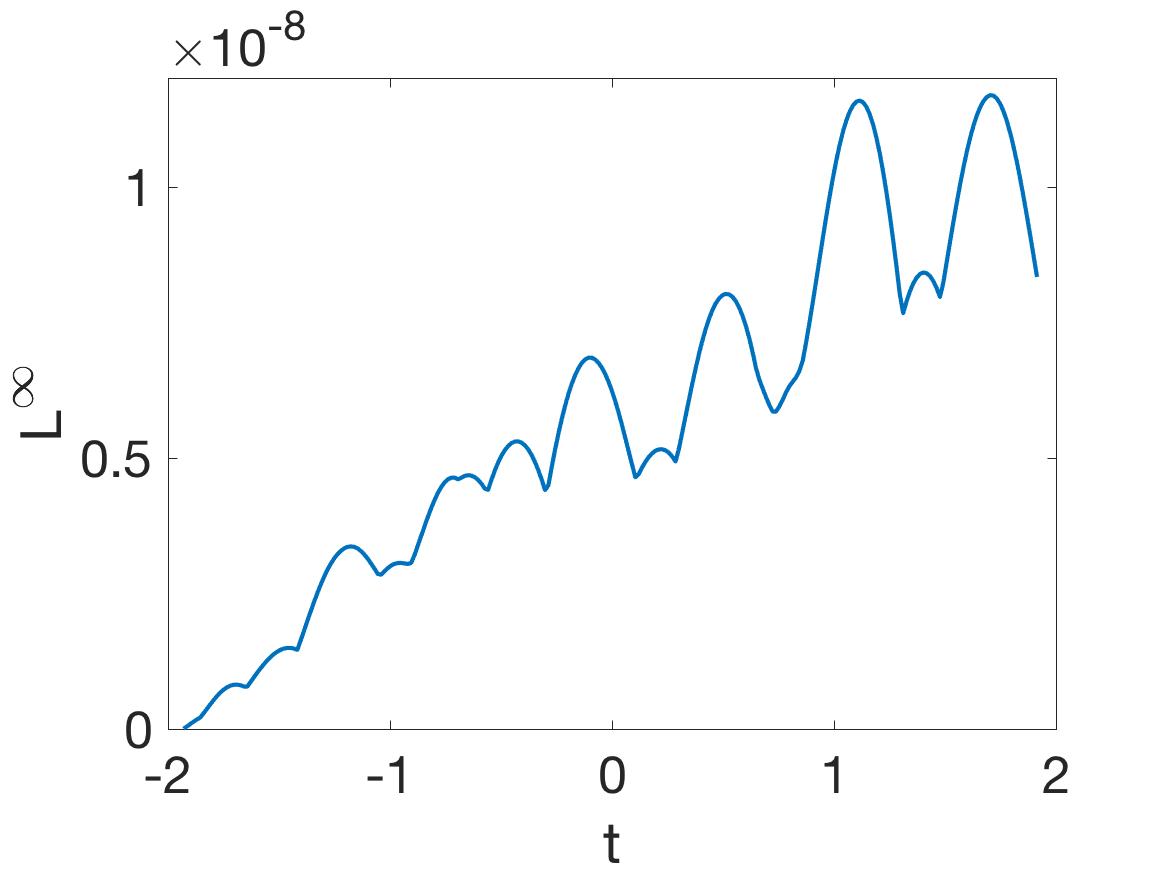}}
\subfloat[$\nu=3$]{\includegraphics[width=.3\textwidth]{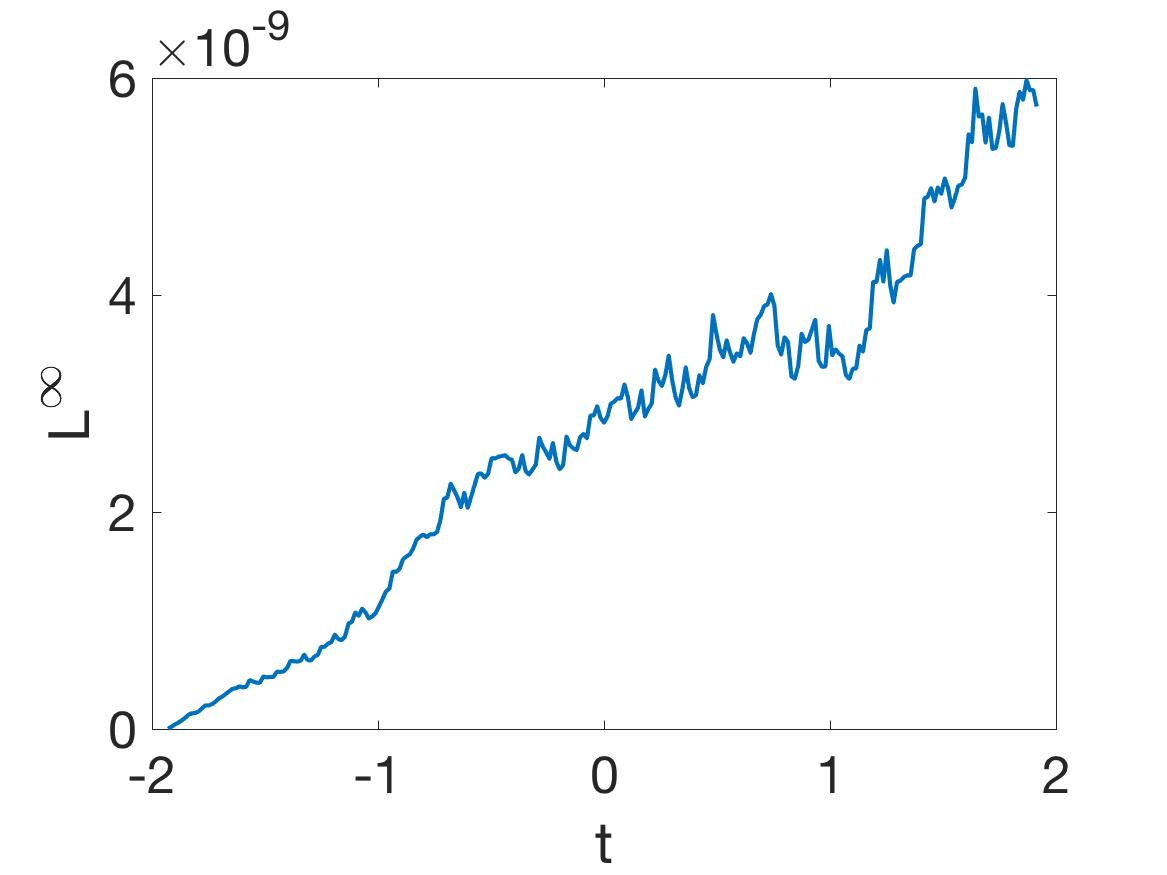}}
%\subfloat[$\nu=4.5$]{\includegraphics[width=.24\textwidth]{Fig13d.jpg}}
%\subfloat[$\nu=6$]{\includegraphics[width=.3\textwidth]{Fig13e.jpg}}
\\
\subfloat[$\nu=6$]{\includegraphics[width=.3\textwidth]{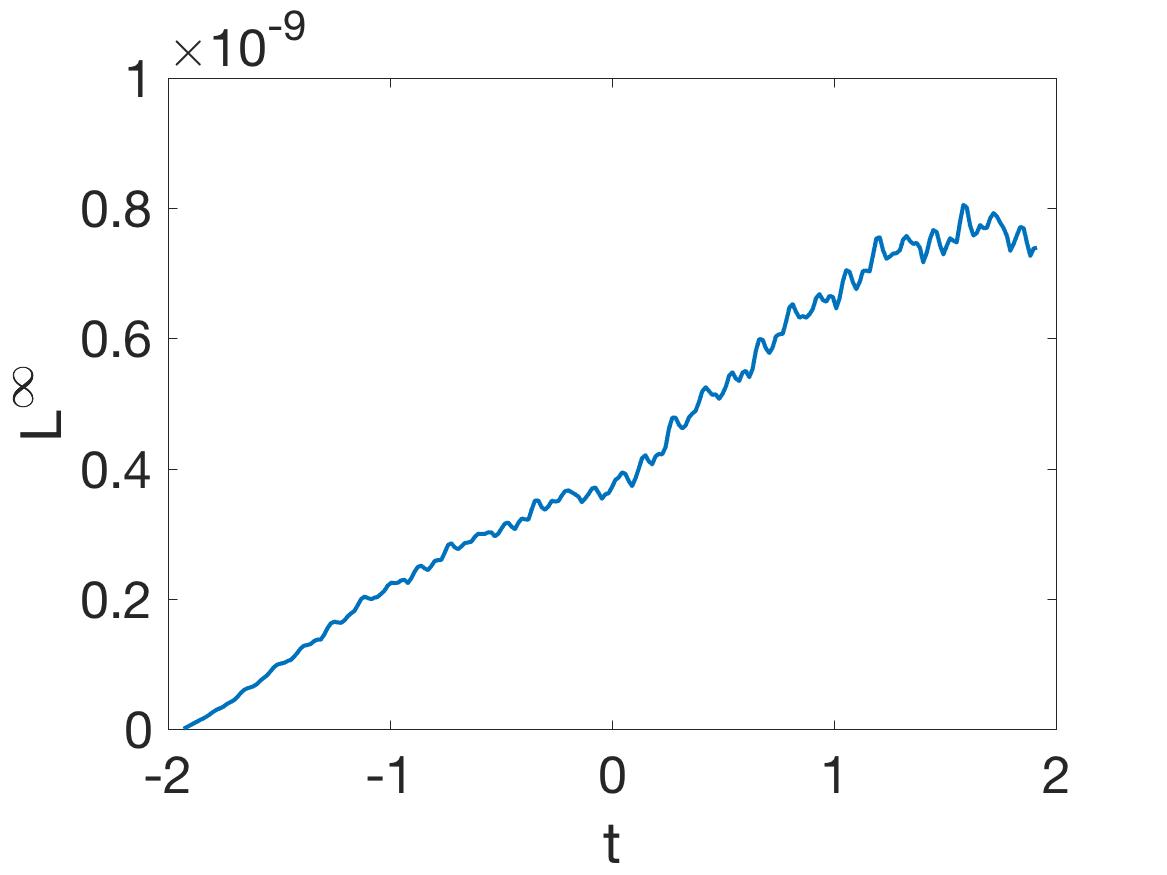}}
\subfloat[$\nu=1.5$]{\includegraphics[width=.3\textwidth]{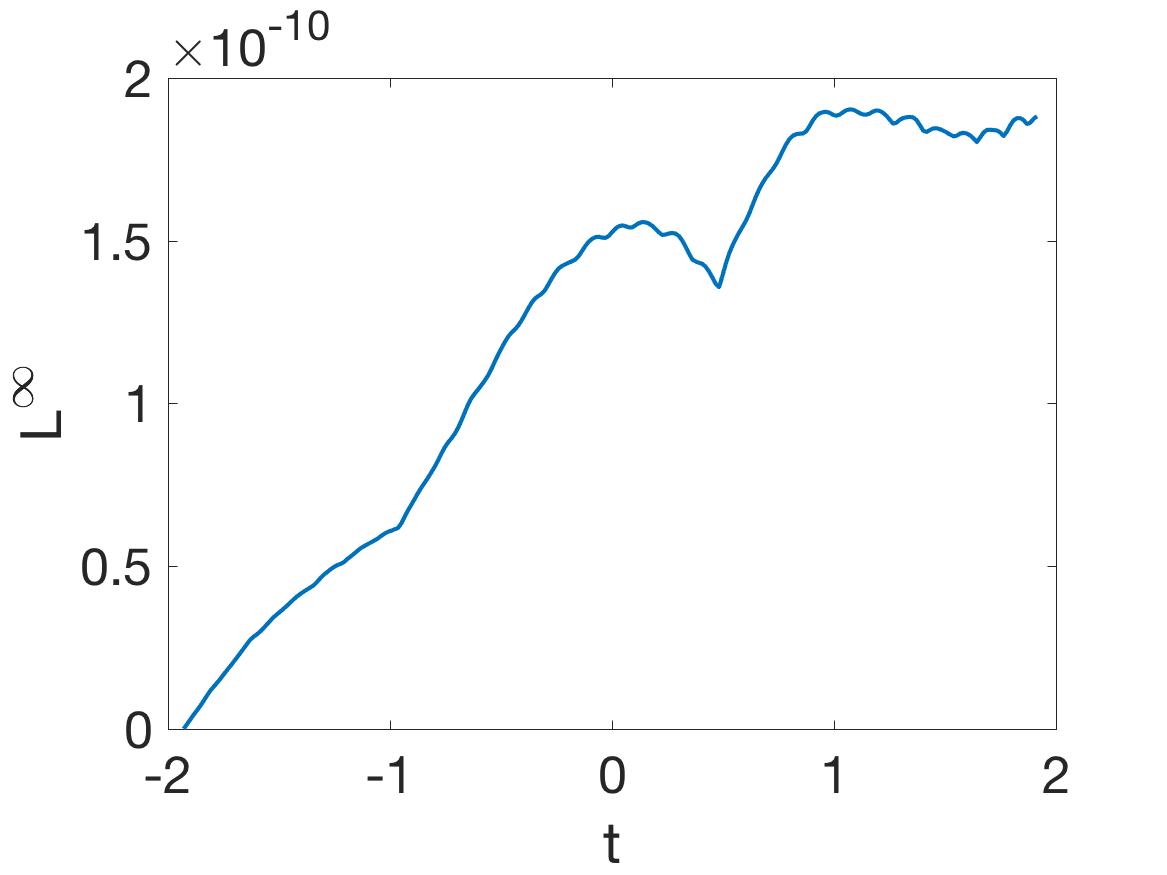}}
\subfloat[$\nu=0$]{\includegraphics[width=.3\textwidth]{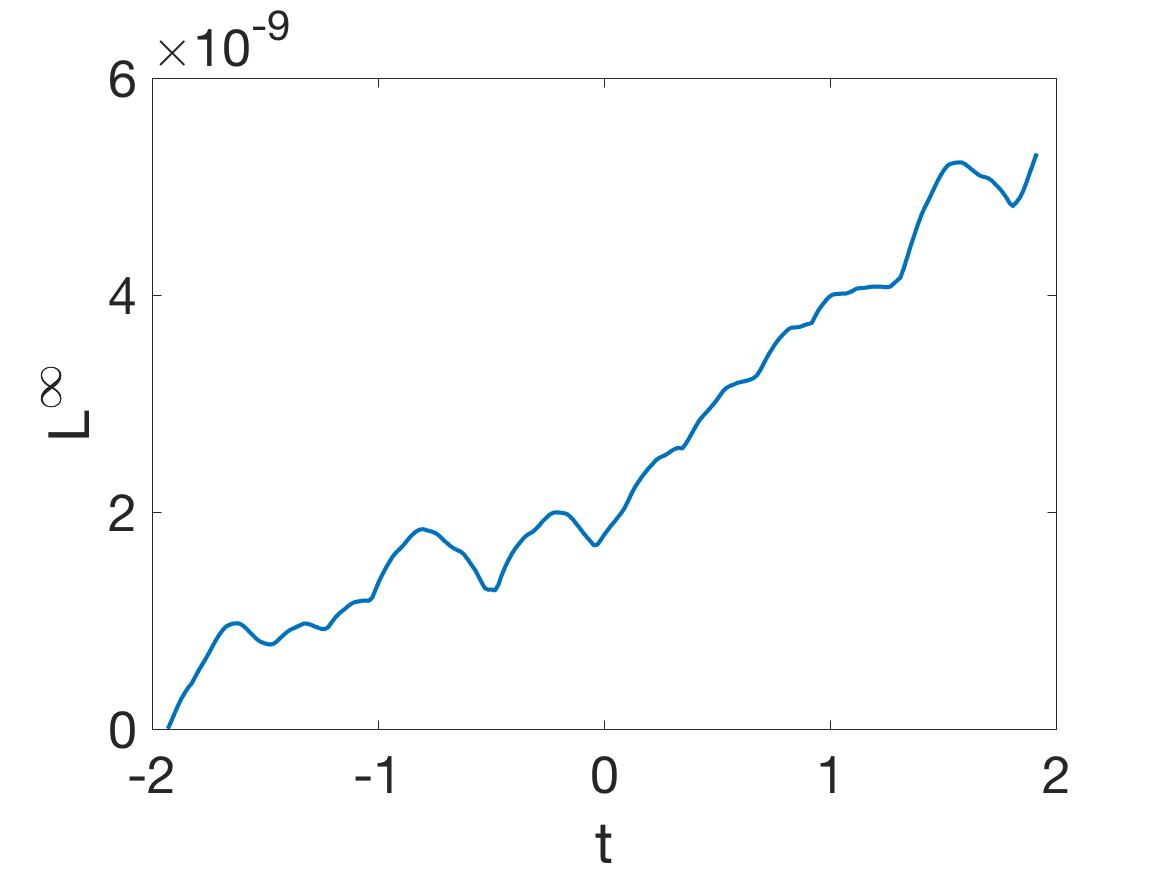}}
\caption{$L^\infty$ error between the solution obtained by the Newton-CG method (as shown in Fig.~\ref{fig:Fig4}) vs the
solution obtained via the ETDRK4  method. The 3 top panels correspond
to the error of the space time periodic solution, while the bottom 3
case examples are examples of the error in the evolution of waveforms
of the stationary modulus branch.}
\label{fig:Fig13}
\end{figure}

%\begin{figure}[]
% Fig 14
%\center
%\subfloat[$\nu=0$]{\includegraphics[width=.24\textwidth]{Fig14a.jpg}}
%\subfloat[$\nu=1.5$]{\includegraphics[width=.24\textwidth]{Fig14b.jpg}}
%\subfloat[$\nu=3$]{\includegraphics[width=.24\textwidth]{Fig14c.jpg}}
%\subfloat[$\nu=4.5$]{\includegraphics[width=.3\textwidth]{Fig14d.jpg}}
%\subfloat[$\nu=6$]{\includegraphics[width=.24\textwidth]{Fig14e.jpg}}
%\caption{$L^\infty$ error between the solution obtained by the Newton-CG method (as shown in Fig.~\ref{fig:Fig5}) vs the
%solution obtained via the ETDRK4  method.}
%\label{fig:Fig14}
%\end{figure}

\section{Conclusions and Future Work}

In the present work, we considered a variety of wave structures existing
in the local NLS equation and extended them to the realm of a generic
nonlocal NLS model. In this vein, we examined the prototypical rogue wave state,
namely the Peregrine soliton, as well as its periodic in time generalization,
the Kuznetsov-Ma (KM) breather. We also looked at other states that are
periodic in both space and time, motivated by the work of~\cite{akh}.
These continuations led to a number of interesting conclusions.
Both the Peregrine soliton and the KM breather were possible to
continue for small values of the nonlocality parameter $\nu$.
This is appealing because it suggests that the
structures are not particular to the integrable limit and can, indeed,
be continued in the non-integrable case.
%It is also intriguing, on the other hand,
%because
Additionally,
as the structures are continued they develop undulations which, in some cases
(e.g., the KM state), suggest connections with other states
that have been recently identified in integrable models, namely
rogue waves mounted on elliptic function, spatially periodic
structures. As regards the doubly-periodic states with periodicities
in both space and time, these were more robustly identified
through our continuation scheme and could, in fact, be continued
up to $\nu=2$ and beyond. However, here too, there exists an
interesting twist, namely the structures beyond a certain degree
of nonlocality lost their temporal periodicity and became
genuinely stationary in their modulus, maintaining only the spatial periodicity.
These spatially periodic states were subsequently continued downward
all the way to the local NLS limit, confirming their cnoidal nature in
the latter and revealing the bifurcation of their spatiotemporally
periodic counterparts.

We believe that these results offer considerable insight on the
potential of the nonlocal model to support states (including rogue wave ones)
with non-vanishing asymptotics, i.e., ones beyond the more
``standard'' solitary wave ones. However, additionally, they also
motivate a number of further questions and inquiries worth
considering in future studies. One of these is whether stationary
elliptic function solutions can be suitably generalized in
analytically available waveforms in the nonlocal case (or whether these
can only be identified in limiting cases such as the ones
of $\nu \rightarrow 0$ and $\nu \rightarrow \infty$
considered herein). Another topic of interest is to explore
systematically continuations of the newly discovered
rogue waves on periodic wave background and explore
how such structures may generalize in the case of the nonlocal
model. Possibly these may be involved in bifurcation phenomena
associated with the states considered here. Another more open
ended challenge is whether rogue-wave-like patterns, such as the
Peregrine soliton or the KM breather, can be continued beyond the intervals
of $\nu$ for which they were found herein in the case of the
nonlocal model.
Lastly, in the present study we focused chiefly on the existence
and in some cases on the bifurcations of the solutions. However,
there are stability tools gradually emerging (such as, e.g., the Floquet
analysis of the KM state and the consideration of the Peregrine as a limiting
case of that calculation~\cite{PRE}) that would be quite relevant to
consider in the present nonlocal setting as well.
Potential progress in any of these directions will be reported in future publications.

\end{document}